\documentclass[amstex]{article}
\usepackage{graphicx}
\usepackage{dcolumn}
\usepackage{amsmath}
\usepackage{amssymb}
\usepackage{amsfonts}
\usepackage{amscd}

\newtheorem{theorem}{Theorem}
\newtheorem{lmm}[theorem]{Lemma}
\newtheorem{cor}[theorem]{Corollary}
\newtheorem{pro}[theorem]{Proposition}
\newtheorem{df}[theorem]{Definition}
\newtheorem{rmk}[theorem]{Remark}

\newcommand\calA{{\cal A}}

\newcommand\calB{{\cal B}}

\newcommand\calH{{\cal H}}
\newcommand\calK{{\cal K}}

\newcommand\calN{{\cal N}}

\newcommand\calS{{\cal S}}

\newcommand\bfZ{\bf Z}

\newcommand\dprime{\prime\prime}
\newcommand\evl{\alpha_t}

\begin{document}
\newpage\thispagestyle{empty}
\begin{center}
{\Huge\bf
Boundedness of Entanglement Entropy,  
\\ 
and
\\
Split Property
\\
of
\\
Quantum Spin Chains}
\\
\bigskip\bigskip
\bigskip\bigskip
{\Large Taku Matsui}
\\
 Graduate School of Mathematics, Kyushu University,
\\
744 Motooka,Nishi-ku Fukuoka 819-0395, JAPAN
\\
 matsui@math.kyushu-u.ac.jp
\\
\bigskip\bigskip
September, 2011
\end{center}
\bigskip\bigskip\bigskip\bigskip
\bigskip\bigskip\bigskip\bigskip
{\bf Abstract:}  
We show the boundedness of entanglement entropy for (bipartite) pure states of quantum spin chains implies
split property of subsystems. As a corollary
the infinite volume ground states for 1-dim spin chains with the spectral gap between the ground state
energy and the rest of spectrum have the split property.  
We see gapless excitation exists  for the spinless Fermion on $\bfZ$ 
if the ground state is  non-trivial  and translationally invariant and the $U(1)$ gauge symmetry is unbroken.
Here we do not assume uniqueness of ground states for all finite volume Hamiltonians.

\newpage
\section{Introduction.}\label{Intro}
\setcounter{theorem}{0}
\setcounter{equation}{0}
In our previous article \cite{Split}, we considered  a relationship between {\it split property} and 
symmetry of of translationally invariant pure states for quantum spin chains on an integer lattice $\bfZ$ .
The {\it split property} is a kind  of statistical independence of left and right semi-infinite subsystems.
More precisely, we say that a state of a quantum spin chain on an integer lattice $\bfZ$
has the split property between left and right semi-infinite subsystems if the state is quasi-equivalent to
a product state of these infinite subsystems.
We have shown that the split property cannot hold for translationally invariant pure states of quantum spin chains 
if the state is SU(2) invariant and the spin $S$ is half-odd integer.
Though this phenomenon looks similar to ground state properties of  antiferromagnetic Heisenberg models  
on the integer lattice $\bfZ$, no direct connection was established there.
The principal purpose of this article is to show that presence of the spectral gap between the ground state
energy and the rest of spectrum implies  the {\it split property} for one-dimensional quantum spin chains. 
We do not assume translational invariance of infinite volume Hamiltonians
and that of states but certain boundedness of the norm of local energy operators. 
\par
The key point of proof of the split property is {\it the boundedness  of entanglement entropy} for
bipartite lattice models. More precisely, we consider pure states of infinite volume systems  and  
the von Neumann entropy of the restriction of states to finite systems in a infinite subsystem, say {\em A}. 
If the entropy is bounded uniformly in the size of the finite systems. we say {\it the entanglement entropy is bounded}. Higher dimensional version of boundedness of the entanglement entropy for bipartite infinite 
quantum systems is the area law of entanglement entropy
The area law of entanglement entropy has been studied in various context of statistical physics and quantum field theory.See \cite{eisert2010} for a overview of the research in this field. 
In Section 2, we will see that pure states satisfying boundedness of entanglement entropy has the split property between two infinite subsystems.
In \cite{Hastings2007},  M.B. Hastings proved the boundedness  of entanglement entropy for ground states 
with a spectral gap and his results implies split property.
M.B. Hastings  assumed that uniqueness of finite volume Hamiltonians in \cite{Hastings2007}.
However, uniqueness condition of finite volume ground states may not be satisfied for AKLT Hamiltonians for which a pure matrix product state is a ground state. 
I.Affleck, T.Kennedy, E.Lieb, H.Tasaki proved that the AKLT model of \cite{AKLT} has a unique infinite volume ground state while the dimension fo the finite volume ground state is four.   
Thus it is natural to expect that ,for any infinite pure ground state with spectral gap,  the split property holds without assuming uniqueness of finite volume ground states.
To prove this, we adapt the proof of the are law of entanglement entropy due to M.B. Hastings 
to an infinite dimensional setting suitably and for that purpose. We find that proof of the factorization lemma 
due to E.Hamza, S.Michalakis, B.Nachtergaele, and R.Sims in  \cite{Nachtergaele2009} is useful.
The improved Lieb-Robinson bound is a crucial mathematical tool for proof of the factorization lemma. 
(See \cite{Koma2006} ,   \cite{LiebRobinson},  \cite{nachtergaele2005},   \cite{Nachtergaele2007a},  \cite{NachtergaeleSims2009}.)
   
\par
As a corollary we will see that
a gapless excitation is present in half-odd integer spin SU(2) invariant quantum spin chains  
and in  $U(1)$ symmetric spinless fermion models on $\bfZ$ provided that the ground state is non-trivial.   
At first sight, our result of gapless excitation in infinite systems may seem to follow from known results of  \cite{AffleckLieb}, \cite{Hal2004} ,\cite{Affleck1997}. 
However,  the previous works is based on the assumption of 
uniqueness of finite volume ground states, 
while we assume only uniqueness of  ground states in infinite systems.
(Our previous result of \cite{Split2} is based on stronger assumption.)
\bigskip 
\noindent
\par
Next we describe results precisely.  We employ the language of operator algebras and most of definitions and notions we use here
can be found  in \cite{BratteliRobinsonI}  and  \cite{BratteliRobinsonII}. 
We describe our results for quantum spin chains on $\bfZ$. Boundedness of entanglement entropy
is a very restrictive condition for higher dimensional translationally invariant systems on $\bfZ^{n}$.  
We denote the $C^{*}$-algebra of (quasi)local observables by $\frak A$. 
 $\frak A$ is the UHF $C^*-$algebra $n^{\infty}$ 
( the $C^{*}$-algebraic completion of the infinite tensor product of n by n matrix algebras ):
$${\frak A} = \overline{\bigotimes_{\bfZ} \: M_{n}({\bf C})}^{C^*} , $$
where $M_{n}({\bf C})$ is the set of all n by n complex matirces.
Each component of the tensor product is specified with a lattice
site $j \in \bfZ$.  ${\frak A}$ is the totality of quasi-local observables.
We denote by $Q^{(j)}$ the element of ${\frak A}$ with $Q$ in
the j th component of the tensor product and the identity in any other
components :
$$  Q^{(j)} = \cdots \otimes 1 \otimes 1 \otimes \underbrace{Q}_{j}  
\otimes 1 \otimes 1 \otimes \cdots$$
For a subset $\Lambda$ of $\bfZ$ , ${\frak A}_{\Lambda}$ is defined as
the $C^*$-subalgebra of ${\frak A}$ generated by elements $Q^{(j)}$ 
with all $j$ in $\Lambda$.
We set
$${\frak A}_{loc} = \cup_{ \Lambda \subset {\bfZ} : | \Lambda | < \infty}
 \:\: {\frak A}_{\Lambda}  $$
where the cardinality of $\Lambda$ is denoted by $|\Lambda |$.
We call an element of ${\frak A}_{loc} $ a local observable
or a strictly local observable. 
\par
By a state $\varphi$ of a quantum spin chain, we mean a normalized positive linear functional
on  $\frak A$ which gives rise to the expectation value of a quantum state.
\par
When $\varphi$ is a state of ${\frak A}$, the restriction of $\varphi$
to ${\frak A}_{\Lambda}$ will be denoted by $\varphi_{\Lambda}$ :
$$\varphi_{\Lambda} = \varphi \vert_{{\frak A}_{\Lambda}} .$$
We set 
$${\frak A}_R =  {{\frak A}}_{[1,\infty)} \: , \:
{\frak A}_L =  {\frak A}_{(-\infty, 0]} \: , \:  \varphi_R=\varphi_{[1,\infty)} \: , \:
\varphi_L = \varphi_{(-\infty, 0]}  \: \: .$$
By $\tau_j$, we denote the automorphism
of $\frak A$ determined by 
$$ \tau_j(Q^{(k)})=Q^{(j+k)}$$ 
for any j and k in $\bfZ$. $\tau_j$ is referred to as the lattice translation of ${\frak A}$.
\par
Given a state $\varphi$ of $\frak A$, we denote the GNS representation of $\frak A$ associated with
$\varphi$ by $\{ \pi_{\varphi}(\frak A) , \Omega_{\varphi}, \frak H_{\varphi} \}$ where
 $\pi_{\varphi}(\cdot)$ is the representation of $\frak A$ on the GNS Hilbert space $ \frak H_{\varphi} $ 
 and $\Omega_{\varphi}$ is the GNS cyclic vector satisfying
 $$\varphi( Q) = \left( \Omega_{\varphi},  \pi_{\varphi}(Q)\Omega_{\varphi}\right) 
 \quad  \: Q \in \frak A .$$
 Let $\pi$ be a representation of $\frak A$ on a Hilbert space.
The von Neumann algebra generated by $\pi({\frak A}_{\Lambda})$
is denoted by ${\frak M}_{\Lambda}$. We set
$${\frak M}_R = {\frak M}_{[1,\infty)} =\pi ({\frak A}_{R})^{\dprime} ,  \quad
{\frak M}_L = {\frak M}_{(-\infty , 0]} =\pi ({\frak A}_{L})^{\dprime} .$$
\bigskip
\par
In terms of the above definitions, we introduce the time evolution of infinite volume systems 
and the ground state in terms of positive linear functionals.
By {\it Interaction} we mean an assignment $\{\Psi (X) \}$ of each finite subset $X$ of 
$\bfZ$ to a selfadjoint operator $\Psi (X)$ in $\frak A_{X}$.
We say that an interaction is of finite range if there exists a positive number $r$ such
that    $\Psi (X) = 0$ if that  the diameter of $X$ is larger than  $r$.
An interaction is translationally invariant if and only if 
$ \tau_{j}(\Psi (X)) = \Psi (X+j)$ for any $X \subset \bfZ$ and for any $j \in \bfZ $.
In what follows, we consider finite range interactions (range = $r$ ), $ \Psi (X) = 0$ if the diameter of $X$ is greater than $r$. If the interaction is not necessarily translationally invariant, 
we assume the following the condition of boundedness : 
\begin{equation}
\sup_{j \in \bfZ} \:\: \sum_{X \ni j}  \:\: \frac{ || \Psi (X) ||}{|X|} < \infty , 
\label{eqn:a1}
\end{equation}
where $|X|$ is the cardinality of $X ( \subset  \bfZ )$.
The infinite volume Hamiltonian $H$ is an infinite sum of  $\{\Psi (X) \}$, 
$$H =   \sum_{X \subset \bfZ} \Psi (X).$$
This sum does not converge in the norm topology, however
the following commutator makes sense:
$$  [ H ,  Q ]  = \lim_{N \to\infty} [ H_{N} , Q] = \sum_{X \subset \bfZ} [ \Psi (X) , Q ]   , \:\:
 \lim_{N \to\infty} e^{it H_{N}} Q  e^{-it H_{N}}   \:\:  Q \in \frak A_{loc} $$
where 
$ H_{N} =  \sum_{X \subset [-N,N]} \Psi (X) $.
\par
Then,the following limit exists for any real $t$:
$$\evl (Q) =  \lim_{N \to \infty} e^{itH_N}  Q e^{-itH_N}$$  
for any element $Q$ of $\frak A$ in the $C^{*}$ norm topology. 
We call $\evl (Q)$ the time evolution of $Q$.
It is known that $\evl (Q)$ as a function of $t$ has an extension to an entire analytic function $\alpha_{z}(Q)$
for any $Q \in \frak A_{loc} $.
\begin{df}
Suppose the time evolution $\evl (Q)$ associated with an interaction
satisfying  (\ref{eqn:a1}) is given.
Let $\varphi$ be a state of $\frak A$. $\varphi$ is a ground state of $H$ if and only if
\begin{equation}
\varphi (Q^{*} [H , Q] )  =  \frac{1}{i} \frac{d}{dt} \varphi (Q^{*} \evl (Q)) \geq 0  
\label{eqn:a2}
\end{equation}
for any $Q$ in $\frak A_{loc}$.
\label{df:GroundState}
\end{df}
Suppose that $\varphi$ is a ground state for $\evl$ . In the GNS representation of
\\
$\{ \pi_{\varphi}(\frak A) , \Omega_{\varphi}, \frak H_{\varphi} \}$, there exists a positive
selfadjoint operator $H_{\varphi} \geq 0$ such that 
$$ e^{itH_{\varphi}} \pi_{\varphi}(Q) e^{-itH_{\varphi}} = \pi_{\varphi}(\evl (Q)), \quad
 e^{itH_{\varphi}} \Omega_{\varphi} =  \Omega_{\varphi} $$
for any $Q$ in $\frak A$.
Roughly speaking, the operator $H_{\varphi}$ is the effective Hamiltonian
 on the physical Hilbert space $ \frak H_{\varphi}$ obtained after regularization
 via subtraction of the vacuum energy.
\par
The spectral gap of  an infinite system is that of $H_{\varphi}$.
Note that, in principle, a different choice of a ground state gives rise
to a different spectrum.
\begin{df}
We say that $H_{\varphi}$ has a spectral gap 
if $0$ is a non-degenerate eigenvalue of $H_{\varphi}$ and for a positive $M >0$,
$H_{\varphi}$ has no spectrum in $(0,M)$,i.e. $Spec (H_{\varphi}) \cap (0,M) = \emptyset $. 
\end{df}
It is easy to see that $H_{\varphi}$ has a spectral gap  if and only if there exists
 a positive constant $M$ such that 
\begin{equation}
\varphi (Q^{*} [H , Q] ) \geq M (\varphi(Q^{*}Q) - |\varphi (Q)|^{2} ) .
\label{eqn:a3}
\end{equation}
\bigskip
\noindent
\par
Now we state our results on split property.
\begin{df}
\label{df:Split}
Let $\varphi$ be a state of $\frak A$.
We say the split property is valid for  $\frak A_{L}$ and $\frak A_{R}$ if
$\varphi$ is quasi-equivalent ti $\psi_{L}\otimes \psi_{R}$
where $\psi_{L}$ is a state of $\frak A_{L}$ and $\psi_{R}$ is that of $\frak A_{R}$.
\end{df}
\begin{df}
\label{df:AreaLaw}
Let $\varphi$ be a state of $\frak A$ and $\rho_{N}$ be the density matrix of $\varphi_{[-N,N]}$.
We consider the entropy $s(\varphi_{[-N,N]}) = - tr_{N} ( \rho_{N} \ln \rho_{N}) = - \varphi (\ln \rho_{N})$
where the trace $tr$ is normalized as $tr(1) =  n^{2N+1}$.
We say the boudedness of entanglement entropy holds for  $\varphi$ if $s(\varphi_{[M,N]})$ is bounded in $N$
,$s(\varphi_{[M,N]}) \leq C$ for any $N$ and $M$ with $M<N$.
\end{df}
\begin{theorem}
\label{th:main1}
\noindent
\newline
Let $\varphi$ be a state of $\frak A$ for which the area law of entanglement entropy holds.
Then the split property is valid  for  $\frak A_{L}$ and $\frak A_{R}$.
\end{theorem}
\begin{cor}
Let $H$ be a finite range Hamiltonian satisfying the boundedness condition (\ref{eqn:a1})
and let $\varphi$ be a ground state of $H$ with spectral gap (\ref{eqn:a3}) .
Then the split property is valid  for  $\frak A_{L}$ and $\frak A_{R}$.
\label{cor:GapSplit1}
\end{cor}
\bigskip
\par
We combine the above results and those of \cite{Split}.
We consider half-odd integer spin $SU(2)$ symmetry of quantum spin chains  and a U(1) symmetry of spinless Fermion,
At this stage we assume translational invariance of Hamiltonians and their ground states.
\par
Let $u(g)$ be the spin $S$ representation of $SU(2)$ and $\gamma_{g}$ be the infinite product type action  
$SU(2)$ on $\frak A$ associated with $u(g)$.
$$(\cdots u(g) \otimes u(g) \otimes \cdots )Q (\cdots u(g) \otimes u(g) \otimes \cdots )^{-1} =\gamma_{g}(Q) , \:\: 
Q \in  \frak A$$
\begin{theorem}
\label{th:SU(2)}
 Consider the quantum spin chain on $\bfZ$ and the spin at each site is a half-odd integer.
Let $H_{S}$ be a translationally invariant , $SU(2)$ gauge invariant  finite range Hamiltonian. 
Suppose that $\varphi$ is a  translationally invariant pure ground state of  $H_{S}$.
Assume that $\varphi$ is $SU(2)$ invariant( $\gamma_{g}$ invariant for any $g$ in $SU(2)$.
Then, there exists gapless excitation in the sense that
$ Spec( H_{\varphi} ) \cap (0,M ) \ne \emptyset $ for any positive $M$. 
\end{theorem}
Next we consider fermions on an integer lattice $\bfZ$. Due to anti-commutativity we impose parity invariance for states, otherwise 
the split property cannot be defined.,
Let $c_{j}^{*}$ and $c_{j} $ be the creation annihilation operators satisfying the standard canonical anti-commutation relations: 
$$\{c_{i} , c_{j}\} =0 ,  \: \{ c_{i}^{*} , c^{*}_{j} \}= 0 , \:   \{ c_{i} ,  c^{*}_{j} \}= \delta_{ij} 1  \quad i,j \in \bfZ$$
By $\frak A^{F}$, we denoted the $C^{*}$-algebra  generated by $c_{i}^{*}$ and $c_{j} $.
$\frak A^{F}$ is referred to as {\it the CAR algebra}.
The sub-algebras $\frak A^{F}_{loc}$, $\frak A^{F}_{\Lambda}$,  $\frak A^{F}_{L}$, $\frak A^{F}_{R}$ of $\frak A^{F}$ are defined as before. 
$\Theta$, $\gamma^{F}_{\theta}$, and $\tau^{F}_{k}$ are automorphisms of the algebra $\frak A^{F}$ 
determined by
$$\Theta (c_{i} )= - c_{i} ,\: \Theta (c^{*}_{i} )= - c^{*}_{i} ,\quad
 \gamma^{F}_{\theta}(c_{i}^{*}) = e^{i\theta} c_{i}^{*} ,\: \quad  \gamma^{F}_{\theta}(c_{i}) = e^{-i\theta} c_{i} ,$$
$$\tau^{F}_{k}(c_{i} ) = c_{i+k} ,\: \tau^{F}_{k}(c_{i}^{*} ) = c_{i+k}^{*} $$
$\gamma^{F}_{\theta}$ (resp. $\tau^{F}_{k}$)) is referred to as the $U(1)$ gauge transformation (resp. translation). $\Theta$ will be called {\em parity}.
\par
Suppose that $\varphi$ is a $\Theta$ invariant state of $\frak A^{F}$.
A product state
$\varphi_{\Lambda}\otimes \varphi_{\Lambda^{c}}$ of $\frak A^{F}$ 
specified with 
$$\varphi_{\Lambda}\otimes_{F} \varphi_{\Lambda^{c}}(Q_{1}Q_{2}) =
\varphi_{\Lambda}(Q_{1})\varphi_{\Lambda^{c}}(Q_{2})  \quad (\: Q_{1}\in \varphi_{\Lambda} , \: Q_{2}\in \varphi_{\Lambda^{c}})$$
can be introduced. The split property for fermion systems may be defined as quasi-equvalence of 
states $\varphi$   and $\varphi_{\Lambda}\otimes_{F} \varphi_{\Lambda^{c}}$.
However for our purpose, the following is convenient.
\begin{theorem}
\label{th:mainFermi1}
\noindent
\newline
Let $\varphi$ be a $\Theta$ invariant pure state of $\frak A^{F}$ for which the area law of entanglement entropy holds.
Then $\pi_{\varphi}(\frak A^{F}_{L})^{\dprime}$ and $\pi_{\varphi}(\frak A^{F}_{R})^{\dprime}$ are type I
von Neumann algebras.
\end{theorem}
We consider Hamiltonians of fermion systems  satisfying 
\begin{eqnarray}
&&H^{F}  = \sum_{j=-\infty}^{\infty}  h_{j}
\nonumber\\
&&  h_{j} \in \frak A^{F}_{[ j-r,j+r]} , \quad \Theta ( h_{j}) =  h_{j} , \quad ||  h_{j} || \leq C
\label{eqn:a4}
\end{eqnarray}
\begin{cor}
Let $H^{F}$ be a finite range Hamiltonian satisfying the boundedness condition (\ref{eqn:a1})
and let $\varphi$ be a ground state of $H^{F}$ with spectral gap (\ref{eqn:a3}) .
Then$\pi_{\varphi}(\frak A^{F}_{L})^{\dprime}$ and $\pi_{\varphi}(\frak A^{F}_{R})^{\dprime}$ are type I
von Neumann algebras.
\label{cor:GapSplitFermi}
\end{cor}
By the standard Fock state we mean the state $\psi_{F}$
 specified by the identity $\psi_{F}(c^{*}_{j}c_{j})=0$ for any $j$ and
  the standard anti-Fock state is the state $\psi_{AF}$ specified by the identity  
  $\psi_{AF}(c_{j}c^{*}_{j})= 0$  for any $j$.
 \begin{theorem}
Consider the spinless Fermion lattice system on  $\bfZ$.
Let $H_{F}$ be a translationally invariant , $U(1)$ gauge invariant  finite range Hamiltonian. 
Suppose that $\varphi$ is a $U(1)$ gauge invariant , translationally invariant pure ground state
of $H_{F}$ and that $\varphi  \ne \psi_{F}$, $\varphi  \ne \psi_{AF}$.
\\
Then, gapless excitation exists between the ground state energy
and the rest of the spectrum of the effective Hamiltonian .
\label{th:mainFermi2}
\end{theorem}

Another application of split property is the distillation of infinitely many copies of the maximally entangled pairs in   
quantum information theory. This was discussed in \cite{{KMSW}}.
We also point out that if the Haag duality holds, Theorem\ref{th:mainFermi2}  and \ref{th:SU(2)} can be shown 
in a different way. The proof of duality in \cite{KMSW2} contains a mistake and we are not able to show the duality
in the general case at the moment. 
\\
\\
\par
In Section\ref{SecArea}, we present our proof of split property assuming boundedness of entanglement entropy
and as an application, we simplify our previous proof that any Frustration Free ground state is a matrix product state 
in Section \ref{FrustrationFree}.
In Section\ref{Factorization}, we will see that the Hastings' factorization lemma implies boundedness of entanglement entropy
in infinite dimensional systems.
In Section \ref{Fermion}, we consider fermionic systems.

\section{Split Property and Entanglement Entropy}\label{SecArea}
\setcounter{theorem}{0}
\setcounter{equation}{0}
In this section we show that the area law of entanglement entropy implies split property. 
\\
First let us recall basic facts of split property or split inclusion of von Neumann algebras.
Let $\frak M_{1}$ and $\frak M_{2}$ be a commuting pair of factors acting on a Hilbert space $\frak H$,
 $\frak M_{1} \subset \frak M_{2}^{\prime}$. We say the inclusion is split
if  there exists an intermediate type I factor $\calN$ such that 
\begin{equation}
\frak M_{1} \subset \calN \subset \frak M_{2}^{\prime} \subset \frak B (\frak H )
\label{eqn:b1}
\end{equation}
The split inclusion is used for analysis of local QFT and of von Neumann algebras and
 some general feature of this concept is investigated for abstract von Neumann alegebras.
by J.von Neumann and later by S.Doplicher and R.Longo in \cite{DoplicherLongo} .
R.Longo used this notion of splitting for his solution to the factorial 
Stone-Weierstrass conjecture in \cite{Longo1}. 
\par
If (\ref{eqn:b1}) is valid, the inclusion of the type I factors 
$\calN =  \frak B (\frak H_{1}) \subset \frak B (\frak H )$tells us factorization of the underlying Hilbert 
spaces and we obtain $\frak H = \frak H_{1} \otimes \frak H_{2}$ and tensor product
\begin{equation}
\frak M_{1}= \tilde{\frak M}_{1}\otimes  1_{\frak H_{2}} 
\subset \frak B (\frak H_{1})\otimes 1_{\frak H_{2}} , \:\:
\frak M_{2}= 1_{\frak H_{1}} \otimes \tilde{\frak M}_{2}  
\subset 1_{\frak H_{1}} \otimes \frak B (\frak H_{2}).
\label{eqn:b2}
\end{equation}
In this sense, the split inclusion is statistical independence of two algebras
$\frak M_{1}$ and $\frak M_{2}$.
\par
If the split  inclusion holds, there exists a normal conditional expectation (partial states) from
the von Neumann algebra $\frak M_{1} \vee \frak M_{2}$   generated by  $\frak M_{1}$ and $\frak M_{2}$  
to  $\frak M_{1}$.
When $\frak M_{2}$ and $\frak M_{1}$ generate  $\frak B (\frak H )$ ,
the split property of the inclusion    $\frak M_{1} \subset \frak M_{2}^{\prime}$
is nothing but  the condition that $\frak M_{1}$ and hence  $\frak M_{2}$  are type I von Neumann
algebras due to the relation $\frak B (\frak H ) =\frak M_{1} \otimes\frak M_{2}$ .
In the present case,  we set 
$\frak M_{1} = \frak M_{\Lambda}=\pi_{\varphi} ( \frak A_{\Lambda})^{\dprime} $, and 
$\frak M_{2} = \frak M_{\Lambda^{c}}=\pi_{\varphi} (\frak  A_{\Lambda^{c}})^{\dprime} $.
By definition, a state $\varphi$  of $\frak A$ satisfies the split property if and only if the following inclusion is split:
$\frak M_{1} \subset \frak M_{2}$
Now we procced to proof of Theorem \ref{th:main1}.
\par
 The state $\varphi$ we consider is pure, and if $\Lambda$ is a finite set of $\bfZ$, there exists the tensor splitting 
of Hilbert spaces;  
\begin{equation}
\frak H_{\varphi} = \frak H_{\Lambda} \otimes \frak H_{\Lambda^{c}}
\label{eqn:b3}
\end{equation}
where the dimension of $ \frak H_{\Lambda}$ is $n^{|\Lambda |}$.
In this splitting, any unit vector $\Omega$ can be written as
\begin{equation}
\Omega \: =  \: \sum_{j=1}^{l}   \: \sqrt{\lambda_{j}}  \xi_{j} \otimes \eta_{j}  
\label{eqn:b4}
\end{equation}
where $0< \lambda_{j+1} \leq \lambda_{j}  \leq \cdots  \leq \lambda_{1} \leq 1, \quad \sum_{j=1}^{l}     \:  \lambda_{j} =1 $
and the orthogonality conditions hold :
$$\xi_{j}\in  \frak H_{\Lambda} , \: \eta_{j} \in \frak H_{\Lambda^{c}} , \quad (\xi_{j} , \xi_{i} )=\delta_{ij} , \: (\eta_{j} , \eta_{i})=\delta_{ij} .$$
Let $\varphi$ be a pure state of $\frak A$ satisfying boundedness of entanglement entropy and 
$\Omega_{\varphi}$ be the GNS cyclic vector
associated with $\varphi$. 
This  factorization (\ref{eqn:b4}) is refered to as Schmidt decomposition.
\\
\\
We set $\Lambda =  [1, N]$ in (\ref{eqn:b4}) and (\ref{eqn:b4} ) is now
\begin{equation}
\Omega_{\varphi} \: = \:  \sum_{j=1}^{l(N)}  \: \sqrt{\lambda^{(N)}_{j}}  \xi^{(N)}_{j} \otimes \eta^{(N)}_{j} .
\label{eqn:b5}
\end{equation} 
Then, in terms of $\lambda^{(N)}_{j}$, the entropy of $s(\varphi_{[1,N]})$ is given by
$$s(\varphi_{[1,N]}) = - \sum_j   \lambda^{(N)}_{j} \ln \lambda^{(N)}_{j} .$$
\begin{lmm}
We set $S = \sup_{N} s(\varphi_{[1,N]})$. Let $k$ be the integer determined by the following conditions:
\begin{equation}
\sum_{j= k+1}^{l(N)}  \: \lambda^{(N)}_{j} < \epsilon,  \quad \quad \sum_{j= k}^{l(N)}  \: \lambda^{(N)}_{j} \geq \epsilon  .
\label{eqn:b6}
\end{equation} 
Then, the following inequalities are valid:
\begin{equation}
 k \leq  \exp (\frac{S}{\epsilon} ) , \quad \exp (- \frac{S}{\epsilon} ) \leq \lambda_{1}.
\label{eqn:b7}
\end{equation} 
\label{lmm:lmm1}
\end{lmm}
{\it Proof} :  We abbreviate $\lambda^{(N)}_{j}$ and $l(N)$ to $\lambda_{j}$ and to $l$. 
As $-\ln \lambda_{j} \leq -\ln \lambda_{j+m}$ for $m >0$, 
we have
$$- \epsilon \ln \lambda_{k} \leq  \sum_{j=k}^{l}   -\lambda_{j} \ln \lambda_{k}  \leq S .$$
Thus $\exp (- \frac{S}{\epsilon} ) \leq \lambda_{k}$. On the other hand,
$ k \lambda_{k} \leq \sum_{j=1}^{k}  \lambda_{j} \leq 1$.
As a consequence, we obtain $ k \leq  \exp (\frac{S}{\epsilon} )$. 
\\ 
\\
\begin{lmm}
Let $\psi_{j, [1,N]}$ be a state of $\frak A_{R}$ which is an extension of the vector state of  $\xi^{(N)}_{j}$
and let  $\varphi_{j, [1,N]^{c}}$ be  a state of $\frak A_{L}$ which is an extension of the vector state of
 $\eta^{(N)}_{j}$).
We can take a sub-sequence $N(m)$ of natural numbers such that we obtain the following (weak*) 
convergence for $j =1,2, \cdots k $: 
$$\psi_{R,j}= \lim_{m} \psi_{j, [1,N(m)]} , \quad \varphi_{L,j}=\lim_{m} \varphi_{j, [1,N(m)]} ,\quad \overline{\lambda}_{j} = \lim_{m} \lambda_{j} .$$
If $\overline{\lambda}_{j} \ne 0$, $\psi_{R,j}$ is quasi-equivalent to $\varphi_{R}$ and $\varphi_{L,j}$ is quasi-equivalent to $\varphi_{L}$
\label{lmm:lmm2}
\end{lmm}
{\it Proof} :@By definition, $\varphi_{R}(Q) = \sum_{j} \lambda_{j} \psi_{j, [1,N]} (Q)$  for $Q \in \frak A_{[1.M]}$ if
$0<M<N$. In paricular $\varphi_{R} \geq  \lambda_{j} \psi_{j, [1,N]}$ if these states are restricted on $\frak A_{[1.M]}$. Then, we take
the weak* limit $N\to \infty$ and we obtain $  \overline{\lambda}_{j} \psi_{R,j} \leq \varphi_{R} $ on $\frak A_{R}$.
As the GNS representation associated with $\varphi_{R}$ is factor,  $\psi_{R,j}$ is quasi-equivalent to $\varphi_{R}$.
The same remark is valid for  $\psi_{L,j}$ is quasi-equivalent to $\varphi_{L}$.
\\
\\
Note that $\frac{1-\epsilon }{k} \leq \overline{\lambda}_{1}$
\\
\\
{\it Proof of Theorem \ref{th:main1}}
\\
We show that $\varphi$ is quasi-equivalent to $\varphi_{L}\otimes \varphi_{R}$.
Because of  Lemma \ref{lmm:lmm2} it suffices to show that $\varphi$ is quasi-equivalent to $\varphi_{L,1}\otimes \varphi_{R,1}$.
We fix a small $\epsilon$ and $k$ as in  Lemma \ref{lmm:lmm1} and set
\begin{equation}
\tilde{\Omega} (N) \: = \:  \sum_{j=1}^{k}  \: \sqrt{\lambda^{(N)}_{j}}  \xi^{(N)}_{j} \otimes \eta^{(N)}_{j} , \quad 
\Omega (N) = \frac{\tilde{\Omega} (N)}{|| \tilde{\Omega} (N)||} .
\label{eqn:b8}
\end{equation} 
Then, 
$$0 < 1- || \tilde{\Omega} (N) ||^{2} < \epsilon , \quad 1- || \tilde{\Omega} (N) || <  \frac{1}{1+ || \tilde{\Omega} (N) || } \epsilon ,
\quad || \tilde{\Omega} (N) - \Omega_{\varphi} ||^{2}  < \epsilon.$$
and 
\begin{eqnarray}
&& || \Omega (N) - \Omega_{\varphi} ||^{2}  = \left( \frac{1}{ || \tilde{\Omega} (N) ||^{2} } - 1\right)
\left( \sum_{j= k+1}^{l(N)}  \: \lambda^{(N)}_{j} \right)    + \left( \sum_{j= k}^{l(N)}  \: \lambda^{(N)}_{j} \right)
\nonumber
\\
&&\leq  \frac{\epsilon }{1-\epsilon} +\epsilon <  3\epsilon .
\label{eqn:b9}
\end{eqnarray}
Let $\omega_{N}$ be the vector state associated with $\Omega (N)$, and let  $\Omega (\infty)$ be
any accumulation point of $\omega_{N}$( in the weak* topology of the state space when we take $N$ to $\infty$. 
Due to (\ref{eqn:b9}),
$$   || \omega_{N} - \varphi || \leq 2\sqrt{3\epsilon }  ,   || \omega_{\infty} - \varphi || \leq 2\sqrt{3\epsilon } $$
which shows that if $ \omega_{\infty}$ is a factor state,  $\omega_{\infty}$ and $\varphi$ are quasi-equivalent.
On the othe hand, by Schwartz inequality, we obtain
\begin{equation}
\omega_{N} \leq  k \sum_{j=1}^{k} \lambda_{j} \psi_{j, [1,N]}\otimes \varphi_{j, [1,N(m)]}, 
\omega_{\infty} \leq  k \sum_{j=1}^{k_{0}} \overline{\lambda}_{j} \psi_{L, j}\otimes \varphi_{R, j} = C \tilde{\varphi}
\label{eqn:b10}
\end{equation}
where $k_{0}$ is the number of $\overline{\lambda}_{j}$ which does not vanish.
$C$ is defined by $ C = k \sum_{j=1}^{k_{0}} \overline{\lambda}_{j}$ and  $\tilde{\varphi}$
is the state of $\frak A$ determined by (\ref{eqn:b10}).
Due to Lemma \ref{lmm:lmm2}, $\psi_{L, j}\otimes \varphi_{R, j}$ are quasi-equivalent to $\varphi_{L}\otimes \varphi_{R}$ 
and hence $\tilde{\varphi}$ is  quasi-equivalent to $\varphi_{L}\otimes \varphi_{R}$ . As a consequence $\tilde{\varphi}$ is a factor state.
The GNS representation associated with $\omega_{\infty}$ is a subrepresentation of that of  $\tilde{\varphi}$ due to (\ref{eqn:b10}).
It turns out that $\omega_{\infty}$  is a factor state quasi-equivalent to $\varphi_{L}\otimes \varphi_{R}$ , which impies split property of
$\varphi$.
{\it End of Proof of Theorem \ref{th:main1}}
\\
\\
\begin{rmk}
In Theorem \ref{th:main1}, we assumed that boundedness of entanglement entropy for our $R$ system.
For pure states without translational invariance, boundedness  of entanglement entropy for the $L$ system
may not follows from that of the $R$ system. A simplest counter example is a pure product states $\varphi = \varphi_{L} \otimes\varphi_{R}$
with $\lim_{N\to \infty} s(\varphi_{[-N,-1]}) = \infty$ . In particular, boundedness of the entanglement entropy for our $R$ system is not a
necessary condition for split property of $\varphi$. On the other hand for states with translational invariance, boundedness 
Theorem \ref{th:main1} can be extended for factor states with an argument similar to that of  Lemma 2 of \cite{Araki1dim}. 
Though the proof is very easy we state it as proposition.
\label{rmk:rmk2.1}
\end{rmk}
\begin{pro}
Let $\varphi$ be a translationally invariant factor state of  a quantum spin chain $\frak A$ on an integer lattice $\bfZ$ and let $s$ be
the mean entropy of  $\varphi$.  Assume that there exists a constant $C$ satisfying
\begin{equation}
| s(\varphi_{[0,n-1]}) - n s | \leq D
\label{eqn:b11}
\end{equation}
 for any $n > 0$. Then, $\varphi$ and $\varphi_{L} \otimes\varphi_{R}$ are quasi-equivalent.
\label{pro:pro2.1}
\end{pro}
{\i Proof:}
We use monotonicity of the relative entropy of a full matrix algebra, say $\calA$. Let $\rho_{1}$ and $\rho_{2}$ be density matrices of states $\varphi_{1}$ and $\varphi_{2}$ and let $s(\varphi_{1}, \varphi_{2})$ be the relative entropy defined by
$s(\varphi_{1}, \varphi_{2})= tr(\rho_{1} \ln \rho_{1} - \ln\rho_{2})$ where we assume that the support of $\rho_{2}$ is smaller than $\rho_{1}$.
For any projection $E$ in $\calA$, due to the monotonicity of $s(\varphi_{1}, \varphi_{2})$, 
\begin{equation}
\varphi_{1}(E)\ln \frac{\varphi_{1}(E)} {\varphi_{2}(E))} + \varphi_{1}(1-E) \ln \frac{\varphi_{1}(1-E)}{\varphi_{2}(1-E)}\leq s(\varphi_{1}, \varphi_{2})
\label{eqn:b12}
\end{equation}
Now we consider a state $\varphi$ of $\frak A$ satisfying the assumption of Proposition {pro:pro2.1} and set
$\varphi_{1} = \varphi_{[-n,n-1]}$ and $\varphi_{2} = \varphi_{[-n,-1]}\otimes \varphi_{[0,n-1]}$. 
By assumption, 
\begin{equation}
 0\leq s(\varphi_{1}, \varphi_{2})= -s(\varphi_{[-n,n-1]}) + 2s( \varphi_{[0,n-1]}) \leq 3D ,
\label{eqn:b13}
\end{equation}
If $\varphi$ and $\varphi_{L} \otimes\varphi_{R}$ are not quasi-equivalent, there exists a projection $E_{\epsilon}$ for a sufficient large $n$
such that $E_{\epsilon}$ is localized in $[-n,n-1]$, and 
$$1-\epsilon \leq \varphi (E) \leq 1 ,\quad 0\leq \varphi_{L}\otimes\varphi_{R}(E) \leq \epsilon .$$
Then, due to (\ref{eqn:b12}), the lefthand side of (\ref{eqn:b13}). Hence the split property holds.
\\
{\it End of Proof of Proposition \ref{pro:pro2.1}}
\section{Frustration Free Ground States}\label{FrustrationFree}
\setcounter{theorem}{0}
\setcounter{equation}{0}
In quantum spin chains, pure states with split property is a generalization of matrix product states
(= finitely correlated states =VBS states) . (c.f.  \cite{AKLT} , \cite{FNW}, \cite{FNW2}  ) Any matrix product state is a frustration free
ground state for a Hamiltonian. More precisely, let $\varphi$ be a translationally invariant matrix product state.
There exists $h \in \frak A_{[0,r]}$ with the following properties: 
$$h= h^{*} \geq 0 ,\quad  \varphi( h ) = \varphi(\tau_{j}(h)) = 0.$$
Set
$$H_{[n.m]}  =   \sum_{j=n}^{m-r} \tau_{j}(h) \in  \frak A_{[n,m]}.$$
Then, $\varphi$ is a ground state of $H_{[n.m]}$ for any $n$, $m$
and the dimension of ground states of finite volume ground Hamiltonians $H_{[n.m]} $ in $\frak A_{[n,m]}$
is finite, bounded uniformly in $n$ and $m$  if $m-n > c_{0}$,  
\begin{equation}
1 \leq dim ker H_{[n.m]} \leq C
\label{eqn:c2}
\end{equation} 
In this section, we consider pure states $\psi$ of  $\frak A_{R}$ satisfying
\begin{equation}
\psi (\tau_{j}(h)) = 0  
\label{eqn:c1a}
\end{equation} 
for any $j \geq 0$ or states $\psi$ of  $\frak A_{[n,m]}$ satisfying
\begin{equation}
\psi (H_{[n.m]}) = 0  
\label{eqn:c1b}
\end{equation} 
The  infinite volume ground state $\varphi$ satisfying the condition (\ref{eqn:c1a}) is called  a frustration free ground state.
The frustration free ground state was called the zero energy state in our previous paper (c.f. \cite{TM1} ) 
but it seems that the word 'frustration free ground state' is frequently used nowadays.
In \cite{TM1} we have shown any frustration free ground state is a matrix product state.  
We present here a simplified proof of the result in  \cite{TM1}.
\\
\par
First we introduce matrix product states.
Let $\calK$ be a n-dimensional Hilbert space.
Suppose that $V$ is an isometry from $\calK$ to 
${\bf C}^d \otimes \calK$.
Consider $E(Q)$ is the linear map from
$M_d({\bf C}) \otimes M_n({\bf C}) $ to $M_n({\bf C}) $
determined by
\begin{equation}
E(Q) = V^* Q V \quad \quad \mbox{ for any $Q$ in 
$M_d({\bf C}) \otimes M_n({\bf C}) $. }
\label{eqn:c3}
\end{equation}
Define
$$ E_Q (R) = E(Q \otimes R) .$$
for $Q$ in $M_d({\bf C}) $ and $R$ in $M_n({\bf C})$.
As $V$ is an isometry, the linear map $E$ and $E_1$ defined above
is unital (= unit preserving $E(1)=1$ , $E_1(1)=1$) CP map . 
Suppose that $\psi$ is a faithful state of $M_n({\bf C})$ satisfying
 the invariance condition below:
\begin{equation}
\psi (R)  = \psi ( E_{1} (R))
\label{eqn:c4}
\end{equation}
where $R$ is any element of $M_n({\bf C})$.
By these data, we can construct a translationally invariant state 
$\varphi$ of the UHF algebra $\calA$ via the following
formula:
\begin{equation}
\varphi ( Q_0^{(j)}  Q_1^{(j+1)} Q_2^{(j+2)} ....Q_l^{(j+l)} )
= \psi (E_{Q_0} \circ E_{Q_1}  ....\circ E_{Q_{j+l}}(1)) .
\label{eqn:c5}
\end{equation}
The state $\varphi$ constructed in this way is called a matrix product state.
\begin{pro}
Suppose that the condition (\ref{eqn:c2}) is valid. 
Let $\varphi$ be a translationally invariant pure ground state .
Then the state $\varphi$ is a matrix product state. 
\label{pro:FrustrationFree}
\end{pro}
We prove Proposition  \ref{pro:FrustrationFree} now.
Let $\rho_{[0,N]}$ be the density matrix of the state $\varphi_{[0,N]}$. As $\varphi_{[0,N]}(H_{[0,N]})=0$
the rank of $\rho_{[0,N]}$ is smaller than $C$ due to the condition  (\ref{eqn:c2}).
This implies the boundedness of the entanglement entropy,  $s(\rho_{[0,N]}) \leq \ln C$.
As a result, $\varphi_{R}$ gives rise to a type I factor representation of $\frak A_{R}$.  
Let $\{ \pi_{0}(\frak A_{R}),\: \frak H_{0}\}$ be the irreducible representation of $\frak A_{R}$ quasi-equivalent 
to the GNS representation associated with $\varphi_{R}$.
There exists the density matrix $\rho_R$ for $\varphi_{R}$:
$$ tr_{\frak H_{0}} ( \rho_{R} Q) = \varphi (Q)   \quad\quad Q \in \frak A_{R}$$
where $ tr_{\frak H_{0}}$ is the trace of $\frak H_{0}$.
\par
We claim that the rank of $\rho_R$ is less than or equal to $C$.
Suppose that
$\rho_R = \sum_{j}  \mu_{j} p_{j}$ where $\mu_{j}$ is a eigenvalue  of  $\rho_R$ satisfying  
$$ 0 < \mu_{j+1}\leq  \mu_{j}  , \quad  \sum_{j=1} \mu_{j} =1 .$$
and $\{ p_{j} \}$ are mutually orthogonal rank one projections associated with eigenvectors $\xi_{j}$. 
Let $\psi_{j}$ be the pure  vector state of  $\frak A_{R}$  associated with the vector $\xi_{j}$.
$\psi_{j}$ also satisfies (\ref{eqn:c1a}) due to the inequality $\mu_{j} \psi_{j} \leq \varphi_{R}$.
$\psi_{j}$ restricted to $\frak A_{[0,N]}$ is a stress free ground state satisfying (\ref{eqn:c1b}).
For each $N > 0 $ there exists a factorization,
$$\xi_{j} =  \sum_{k}  \lambda_{k}(N,j) \eta_{[0,N]}(k,j) \otimes \eta_{[N ,\infty)} (k,j)$$
where $\eta_{[0,N]}(k,j)$ are mutually orthogonal unit vectors of the spin chain on $[0,N]$ and $\eta_{[N ,\infty)} (k,j)$ are those on $[N ,\infty)$
and  $\lambda_{k}(N,j)$ is a positive number  satisfying 
\begin{equation}
0 \leq \lambda_{k+1}(N,j) \leq \lambda_{k}(N,j) , \quad \sum_{k}  \lambda_{k}(N,j ) = 1
\label{eqn:c6}
\end{equation}
We have
\begin{equation}
\psi_{j} (Q) =  \sum_{k=1}^{C}  \lambda_{k}(N,j)  (\eta_{[0,N]}(k,j), Q\eta_{[0,N]}(k,j)) \quad\quad Q \in \frak A_{[0,N]} ,
\label{eqn:c7}
\end{equation}
which shows that vector states associated with $\eta_{[0,N]}(k,j)$ and $\eta_{[N ,\infty)} (k,j)$ are stress free ground states as well.
It turns out that the number of the summand in (\ref{eqn:c6}) cannot exceed the dimension of finite volume stress free ground states,
and $\frac{1}{C} \leq   \lambda_{1}(N,j )$. 
\begin{lmm}
Any weak* accumulation point of the vector state  associated with $\eta_{[0,N]}(1,j)$ (when $N \to \infty$) is $\psi_{j}$.
\label{lmm:lmm3.2} 
\end{lmm}
{\it Proof:} Let $\overline{\eta}_{j}$ be any weak* accumulation point. Due to (\ref{eqn:c7}) we have $ \frac{1}{C} \overline{\eta}_{j} \leq \psi_{j}$.
As $\psi_{j}$ is a pure state of $\frak A_{R}$ we conclude that $\overline{\eta}_{j} = \psi_{j}$
{\it End of Proof.}
\bigskip
\bigskip
\newline
The following lemma shows that $\eta_{[0,M]}(1,i)$ are asymptotically orthogonal.
\begin{lmm} 
For any $\epsilon$, there exists $N$ such that  for any $M$ with $M \geq N$
\begin{equation}
 | (  \eta_{[0,M]}(1,i) , \eta_{[0,M]}(1,j) ) | \leq \epsilon 
\label{eqn:c8}
\end{equation}
if  $i\ne j \leq C$.
\label{lmm:lmm3.3}
\end{lmm}
{\it Proof:}  As $p_{j}$ is in $\frak M_{R}$, there exists a projection $E_{j} \in \pi_{0}( \frak A_{[0,N(1)]})$ such that
$$  0\leq 1 - \psi_{j}(E_{j}) < \epsilon  \: , \quad 0 \leq \psi_{i}(E_{j}) \leq \epsilon  \:\: (i \ne j ).$$
We have $N(2)$ such that the following is valid for any $M > N(2)$: 
$$ 0\leq (  \eta_{[0,M]}(1,j) , (1- E_{j}) \eta_{[0,M]}(1,j) ) \epsilon ,\quad 0 \leq (  \eta_{[0,M]}(1,i) , A_{j}\eta_{[0,M]}(1,i) )  < \epsilon$$
for $i \ne j \leq C $. Then, 
\begin{eqnarray*}
&& | (  \eta_{[0,M]}(1,i) , \eta_{[0,M]}(1,j) ) | \leq   |( \eta_{[0,M]}(1,i) , E_{j}\eta_{[0,M]}(1,j) ) | + \sqrt{\epsilon}
\\
&&=   (E_{j} \eta_{[0,M]}(1,i) , \eta_{[0,M]}(1,j) ) | + \sqrt{\epsilon}
\leq (  \eta_{[0,M]}(1,i) , E_{j}\eta_{[0,M]}(1,i) )^{1/2} + \sqrt{\epsilon} \leq  2 \sqrt{\epsilon}. 
\end{eqnarray*}
As the above $\epsilon$ is arbitrary, we obtain $N$ satisfying (\ref{eqn:c8}).
{\it End of Proof.}
\bigskip
\bigskip
\newline
\begin{lmm} 
Suppose that $\{ x_{1}, \cdots , x_{L}\}$ are unit vectors in a Hilbert space and 
assume that 
\begin{equation}
| ( x_{i} , x_{j} )| < \epsilon \quad\quad\mbox{ for $i\ne j$ .}
\label{eqn:c9}
\end{equation}
If $0 < \epsilon < \frac{1}{L}$, $\{ x_{1}, \cdots , x_{L}\}$ are linearly independent.
\label{lmm:lmm3.4}
\end{lmm}
{\it Proof:}
We consider complex numbers $c_{j}$ satisfying $\sum_{j=1}^{L} c_{j}  x_{j} = 0$. This equation is written in a matrix form:
$$ (1 + B )c = 0, \quad  B_{ij} = ( x_{i} , x_{j} ) , \quad c = \left(\begin{array}{c}c_{1} \\ \cdot \\ \cdot \\ c_{L}\end{array}\right)$$
where $B_{ij}$ is the $(i,j)$ component of the hermitian matrix $B$.
Due to the condition  (\ref{eqn:c9}) the operator norm of $B$ is less than $(n-1) \epsilon$ and  $1 + B$ is strictly positive matrix.
Hence, $c=0$. {\it End of Proof.}
\\
\\
Lemma \ref{lmm:lmm3.3} and Lemma \ref{lmm:lmm3.4} tell us that  $\{ \eta_{[0,M]}(1,i) \}$ are linearly independent and
the number of these vectors $\eta_{[0,M]}(1,i)$ cannot exceed the dimension of stress free ground states of $H_{[0,M]}$ .
As a consequence, the rank of the density matrix $\rho_{R}$ is finite. 

The rest of proof of Proposition \ref{pro:FrustrationFree} is easy. As the state $\varphi_{R}$ is of type I the GNS reprsentation
gives rise to a shift of $\calB (\frak H_{0})$ associated with the lattice translation $\tau_{1}$. 
(c.f. \cite{Powers}) 
There exists a representation of the Cuntz algebra $O_{d}$ with standard generators $S_{j}$ which implements the shift $\tau_{1}$ :    
$$ \sum_{j=1}^{d} S_{j} \pi_{0}(Q)S_{j}^{*} =  \pi_{0}(\tau_{1}(Q)) \quad\quad Q \in \frak A_{R}$$
Let $P$ be the support projection of $\varphi_{R}$ for $\frak M_{R}$. The range of $P$ (in $\frak H_{0}$) is finite dimensional
and set $\calK = P\frak H_{0}$  $V_{j} = S_{j}^{*}P= PS_{j}^{*}P$ and let $\psi$ be the restriction of  
$\varphi$ to $\calB (\calK ) = P\calB (\frak H_{0})P$ . 
$V$ is an isometry from $\calK$ to  ${\bf C}^d \otimes \calK$ determined by
$Vx = (PS_{1}^{*}Px, \cdots PS_{j}^{*}Px\cdots PS_{d}^{*}P )$.
With these staffs, it is straight forward to see that $\varphi$ is the matrix product state associated with $\{ V, \calK , \psi \}$.

\section{Factorization Lemma of M.Hastings}\label{Factorization}
\setcounter{theorem}{0}
\setcounter{equation}{0}
In  \cite{Hastings2007} M.Hastings proved boundedness of entangled entropy for gapped ground states.
What M.Hastings proved was estimates of entropy uniformly in sizes of finite volume ground states,
which is not exactly same as what we need for split property. We explain here a minor technical difference. 
The proof below is essentially due to M.Hastings.
\\
\\
Let $H$ be a finite range Hamiltonian with the boundedness condition (\ref{eqn:a1}) and
$\evl$ be the associated time evolution.
Suppose $\varphi$ is a ground state of $H$ satisfying the gap condition (\ref{eqn:a3}).
On $\frak H_{\varphi}$ there exists a positive self-adjoint operator  $H_{\varphi}$ satisfying
$e^{it H_{\varphi}} \pi_{\varphi}(Q)e^{-it H_{\varphi}} =  \pi_{\varphi}(\evl (Q))$
and $H_{\varphi}\Omega_{\varphi}=0$.
We set  $s_{n} =  \sup \{ s(\varphi_{[0,j]} )  \: |\:   0 \leq j \leq n \}$ and our aim is to show $\lim_{n} s_{n} <\infty$.
\\
\\
Let $P_{0}$ be the rank one projection $| \Omega_{\varphi} ><\Omega_{\varphi}|$ to the ground state vector $\Omega_{\varphi}$.
The following lemma is refered to as Hastings' Factorization Lemma
\begin{lmm}
Suppose $\frak H_{\varphi}$ has a spectral gap (specified in (\ref{eqn:a3}))    
For any $n$ and $l (< n/8)$ there exist positive contants $C_{1}$, $C_{2}$, 
\begin{equation}
|| \: O_{B}(n,l)O_{R}(n,l)O_{L}(n,l) - P_{0} \: || \leq C_{1}\exp ( -C_{2} l ) \equiv \epsilon (l) .
\label{eqn:d1}
\end{equation}
where $O_{L}(n,l)$, $O_{R}(n,l)$ are projections and $O_{B}(n,l)$ is a positive selfadjoint operator 
satisfying 
\begin{eqnarray}
&&O_{R}(n,l) \in \pi_{\varphi}(\frak A_{[0,n-1]}), \quad  O_{L}(n,l) \in    \pi_{\varphi}(\frak A_{[0,n-1]^{c}})^{\dprime},
\label{eqn:d1x1}
\\
&&O_{B}(n,l) \in \pi_{\varphi}( \frak A_{(-l,l)\cup (n-l,n+l)}) , 
\label{eqn:d2}
\\
&& 0\leq  0\leq O_{B}(n,l)\leq 1 
\label{eqn:d3}
\end{eqnarray}
\label{lmm:lmm4.1}
\end{lmm}
Due to (\ref{eqn:d1}) $|| [O_{B}(n,l)\: , \: O_{R}(n,l)O_{L}(n,l)\: ] || \leq 2 \epsilon (l)$ so by changing constants
we may assume 
\begin{equation}
|| [O_{B}(n,l) , O_{R}(n,l)O_{L}(n,l)] || \leq \epsilon (l) .
\label{eqn:d4}
\end{equation}
By (\ref{eqn:d1}), (\ref{eqn:d2}) and (\ref{eqn:d3}),
\begin{equation}
 1-\epsilon (l) \leq \varphi (O_{R}(n,l)O_{L}(n,l) ) , \quad  1-\epsilon (l) \leq \varphi (O_{B}(n,l)) .
\label{eqn:d5}
\end{equation}
\\
\\
Boundedness of entanglement entropy follows from Hastings' factorization lemma.
Detail of construction of operators $O_{R}(n,l)$, $O_{L}(n,l)$ and $O_{B}(n,l)$ is not used 
in  the next step of proof.
Here we explain an itinerary from  Hastings' Factorization Lemma to boundedness of entanglement entropy.
\begin{pro}
Suppose that  there exist projections $O_{R}(n,l)$, $O_{L}(n,l)$ and $O_{B}(n,l)$ satisfying 
(\ref{eqn:d1}), (\ref{eqn:d1x1}), (\ref{eqn:d2}) and (\ref{eqn:d3}).  
Then, the entanglement entropy is  bounded:  $\sup_{n} s(\varphi_{[1,n]} < \infty$
\label{pro:pro4.2}
\end{pro}
\bigskip
\par
We set $[0,n]^{c}=(-\infty , -1]\cup [n+1,\infty)$, $\varphi_{R,n} = \varphi_{[0,n]} $ and  
$\varphi_{L,n} = \varphi_{[0,n]^{c}}$ . 
The density matrix of $\varphi_{R,n} $ ( resp. $\varphi_{L,n}$) will be denoted by $\rho_{R, n}$
(resp. $\rho_{L,n}$). 
The Schmidt decomposition (\ref{eqn:b4}) shows that the entanglement entropy and the rank of $\rho_{L,n}$ are equal to those of $\rho_{R,n}$.
\\
\begin{lmm}
We define $p$ via the following equation:
\begin{equation}
p = (\Omega_{\varphi} , \rho_{L,n}\otimes \rho_{R,n}\Omega_{\varphi}) = \varphi (\rho_{L,n}\otimes \rho_{R,n})
\label{eqn:d6}
\end{equation}
where by abuse of notation we use $\varphi$ for the normal extension of $\varphi$ to 
$\frak M = \pi_{\varphi}(\frak A )$.
Then,
\begin{equation}
s(\varphi_{[0,n]}) \leq C_{2}\ln (2C_{1}^{2}/p ) \ln 4d  + F
\label{eqn:d7}
\end{equation}
where $F= (C_{2}+4)\ln 4d +1 + \ln (d^{8}-1)+\ln (C_{2}/2 +1)$.
\label{lmm:lmm4.2}
\end{lmm}
To show Lemma \ref{lmm:lmm4.2} we use the following min-max principle.This should be known, though, as we are not aware
of any suitable reference, we include its proof here. 
\begin{lmm}
Let $\rho$ be a hermitian matrix acting on a $N$ dimensional space and let $\rho_{k}$ be the eigenvalue of $\rho$ 
satisfying $\rho_{1} \geq \rho_{2} \cdots \rho_{k} \geq \rho_{k+1} \cdots \geq \rho_{N}$.
Set
$$\mu_{k} = \sup\{ tr(\rho E ) \: | \: E^{*}=E=E^{2} ,tr(E) = k\} .$$
i.e. the supremum is taken among projections $E$ with rank $k$. 
Then,
$$\mu_{k} = \sum_{i=1}^{k}  \rho_{i}$$
\label{lmm:lmm4.3}
\end{lmm}
{\it Proof of Lemma \ref{lmm:lmm4.3}:} 
\\
Let ${\cal V}_{k}$ be a $k$ dimensional subspace. There exists a vector $\xi \in {\cal V}_{k}$ such that
$(\rho \xi , \xi )\leq \rho_{k}$. This is because the $N-k +1$ dimensional subspace $\calS$ spanned by eigenvectors with eigenvalues
$ \rho_{k}, \rho_{k+1} \cdots \rho_{N}$ has non-trivial intersection with ${\cal V}_{k}$ . ( ${\cal V}_{k}\cap \calS = \{0\}$ implies that
the dimension of the total vector space is $N+1$.) Now we show our claim by induction of the dimension $k$.
By definition, $\mu_{k} \geq \sum_{i=1}^{k}  \rho_{k}$ and we assume that $\mu_{k-1} = \sum_{i=1}^{k-1}  \rho_{i}$. Let $E_{k}$ be 
a  rank $k$ projection and take a unit vector $\xi$ in the range of $E_{k}$ such that $(\rho \xi , \xi )\leq \rho_{k}$. 
For the projection $F$ to the orthogonal complement of $\xi$ in the range of $E_{k}$, we have 
$ tr(\rho F )\leq  \sum_{i=1}^{k-1}  \rho_{i}$ and as a consequence, we obtain  
$$tr(\rho E_{k}) = tr(\rho F )+ (\rho \xi , \xi ) \leq  \sum_{i=1}^{k}\rho_{i}.$$ 
{\it End of Proof of Lemma \ref{lmm:lmm4.3}} 
\bigskip
\bigskip
\newline
\noindent
Let $\xi$ be a vector in a tensor product of Hilbert spaces $\frak H_{1} \otimes  \frak H_{2}$ 
and $\{ \Psi_{k} \}$ (resp.  $\{ \Phi_{l} \}$ ) be a CONS of  $\frak H_{1}$ (resp. $\frak H_{2}$).
Then $\xi$ can be written as
$$\xi = \sum_{l,k} c_{kl} \Psi_{k} \otimes \Phi_{l}.$$
We say $\xi$ has the Schmidt rank $K$ if the rank of the matrix $C$ with entries $c_{kl}$ is $K$.  
The Schmidt rank of $\xi$ can be determined independent of choice of CONS of $\frak H_{1}$ and $\frak H_{2}$.
For a vector $\xi$ with the Schmidt rank $K$ the Schmidt decomposition is equivalent to the existence of 
 CONS $\{ \Psi_{k} \}$ of  $\frak H_{1}$ and $\{ \Phi_{k} \}$ of $\frak H_{2}$ such that
$$\xi = \sum_{k=1}^{K} c_{k} \Psi_{k} \otimes \Phi_{k} , \quad c_{k}\geq 0 , \quad \sum_{k=1}^{K} c_{k}^{2} = ||\xi || .$$
We say that a density matrix $\rho$ on $\frak H_{1} \otimes  \frak H_{2}$
has the Schmidt rank at most $K$ if the Schmidt rank of any eigenvector of $\rho$ is less than or equal to $K$. 
\bigskip
\bigskip
\newline
\noindent
{\it Proof of Lemma \ref{lmm:lmm4.2}:} 
\\
Set $\rho(n,l) = \rho_{L,n}\otimes \rho_{R,n}$
and
$$\tilde{\rho}(n,l) =  O_{B}(n,l)O_{R}(n,l)O_{L}(n,l) \: \rho(n,l) \: O_{R}(n,l)O_{L}(n,l)O_{B}(n,l) .$$
Then, for the norm $||A||_{\varphi} =||AP_{0}||_{\varphi}= \varphi (A^{*}A)^{1/2}$, we obtain
$$||\rho(n,l)^{1/2} ||_{\varphi}  \leq  || \rho(n,l)^{1/2} \: O_{R}(n,l)O_{L}(n,l) O_{B}(n,l) ||_{\varphi}  +   || B||_{\varphi} $$
where $B =  \rho(n,l)^{1/2}) \{ P_{0} - O_{R}(n,l)O_{L}(n,l)O_{B}(n,l)\}$.
As $||B|| \leq \epsilon$ we have
\begin{equation}
\sqrt{p} - \epsilon   \leq  \varphi (\tilde{\rho}(n,l))^{1/2} .
\label{eqn:d8}
\end{equation}
Now we claim
\begin{equation}
1- 2\frac{\epsilon^{2}(l)}{p} \leq  \frac{\varphi (\tilde{\rho}(n,l))}{tr((\tilde{\rho}(n,l))} .
\label{eqn:d9}
\end{equation}
If $\epsilon (l) \geq \sqrt{p}$, the left-hand side of  (\ref{eqn:d9}) is negative. We may assume  $0 \leq \sqrt{p} -\epsilon (l)$. 
Then,
\begin{eqnarray}
&&1- \frac{\varphi (\tilde{\rho}(n,l))}{tr(((1-P_{0})+P_{0})(\tilde{\rho}(n,l))} 
=  \frac{tr((1-P_{0})(\tilde{\rho}(n,l))}{\varphi (\tilde{\rho}(n,l)) + tr((1-P_{0})(\tilde{\rho}(n,l))} 
\nonumber
\\
&&\leq \frac{\epsilon^{2}(l)}{\epsilon^{2}(l) + \varphi (\tilde{\rho}(n,l))}
\leq \frac{\epsilon^{2}(l)}{\epsilon^{2}(l)+ (\sqrt{p} - \epsilon (l))^{2}} = \frac{\epsilon^{2}(l)}{2(\epsilon (l) -1/2 \sqrt{p} )^{2}+ 1/2\cdot p}
\nonumber
\\
&&\leq \frac{\epsilon^{2}(l)}{1/2\cdot p}.
\label{eqn:d10}
\end{eqnarray}
Next we consider the Schmidt decoposition of the ground state vector $\Omega_{\varphi}$ for $\Lambda = [0, n-1]$ in (\ref{eqn:b4})
$$\Omega_{\varphi} \: =  \: \sum_{j=1}^{l}   \: \sqrt{\lambda_{j}}  \xi_{j} \otimes \eta_{j}, \quad
0< \lambda_{j+1} \leq \lambda_{j}  \leq \cdots  \leq \lambda_{1} \leq 1, \quad \sum_{j=1}^{l}     \:  \lambda_{j} =1 $$
where $\{\xi_{j}\}$ is an orthogonal system of $\frak H_{[0, n-1]}$ and
$\{\eta_{j}\}$  is that of $\frak H_{(-\infty,-1]\cup [n,\infty)}$.
We claim that 
\begin{equation}
\sum_{d^{8l-4}+1\leq j} \lambda_{j} \leq \frac{2 \epsilon^{2}(l)}{p}.
\label{eqn:d11}
\end{equation}
Consider the density matrix $\rho$ defined by
$$\rho = \frac{\tilde{\rho}(n,l)}{tr((\tilde{\rho}(n,l))} = \sum_{j=1} \: \mu_{j} \: |x_{j}><x_{j}| $$
where $x_{j}$ is an eigenvector for the eigenvalue $\mu_{j}$ and we assume $\mu_{j+1}\leq\mu_{j}$.
As the Schmidt rank of $\rho_{1}\otimes\rho_{2}$ is one, and as $O_{B}(n,l)$ is in the $d^{8l-4}$ dimensional space 
$\frak A_{(-l,l)\cup (n-l,n+l)}$, the Schmidt rank of $x_{j}$ is at most $d^{8l-4}$. Set $M=8l-4$.
We  may express $x_{j}$ in a linear combination of $\xi_{j} \otimes \eta_{j}$ as follows:  
$$x_{j} = \sum_{kl} \: c^{j}_{kl} \:  \xi_{k} \otimes \eta_{l} \: , \quad \sum_{kl} \: |c_{kl}(j)|^{2} =1 .$$
Then,
\begin{equation}
(\Omega_{\varphi}, \rho \Omega_{\varphi}) = \sum_{j}  \mu_{j} \: |\: \sum_{k} \sqrt{\lambda_{k}} c_{kk}(j) \: |^{2} .
\label{eqn:d12}
\end{equation}
Let $\Lambda$ and $C(j)$ be matrices with entries defined by
$$ \Lambda_{kl} = \delta_{kl} \lambda_{k} , \quad C_{kl}(j) = c_{kl}(j) .$$
$\Lambda$ is a non-negative matrtix with $tr(\Lambda )= 1$ and the rank of $C(j)$ is at most $d^M$ and $tr((C(j)^{*} C(j))=1$.
By the support projection $E(j)$ of $C(j)$ we mean the minimal projection satisfying $E(j)C(j) = C(j)$, and (\ref{eqn:d12}) is 
written as 
\begin{eqnarray*}
&&(\Omega_{\varphi}, \rho \Omega_{\varphi}) =  \sum_{j}  \mu_{j} \: | tr(\Lambda^{1/2} E_{j}C(j)) |^{2}
\\
&\leq &\sum_{j}  \mu_{j}  tr(\Lambda^{1/2} E_{j} \Lambda^{1/2})  tr((C(j)^{*}C(j)) = \sum_{j}  \mu_{j}  tr(\Lambda E_{j} )
\end{eqnarray*}
As the rank of $E(j)$ is at most $d^M$ Lemma \ref{lmm:lmm4.3} implies
$ tr(\Lambda E_{j} ) \leq \sum_{k=1}^{d^M} \lambda_{k}$. 
Thus we have
$$(\Omega_{\varphi}, \rho \Omega_{\varphi}) \leq \sum_{k=1}^{d^M} \lambda_{k} $$
which shows (\ref{eqn:d11}).

Next we give the estimate of the entropy (\ref{eqn:d7}). We use
\begin{equation}
\sum_{j=1}^{K-1} - x_{j} \ln x_{j} \leq  \left(\sum_{j=1}^{K-1 } x_{j} \right) \ln K - \left(\sum_{j=1}^{K-1 } x_{j} \right) \left\{  \ln (\sum_{j=1}^{K-1} x_{j}) \right\}
\label{eqn:d13}
\end{equation}
 for any non-increasing sequence of positive numbers $x_{j}$.
Assuming the conditions (i) $0\leq a_{j+1}\leq a_{j} \leq \cdots \leq a_{1}\leq 1$ and (ii)
$\sum_{j \geq k}  x_{j} \leq a_{k}$ for all $k =1,2,\cdots$  we have the following bound:
\begin{equation}
\sum_{j=1}^{\infty}  - x_{j} \ln x_{j} \leq \sum_{k=1}  - (a_{k}-a_{k+1})\ln (a_{k}-a_{k+1}) . 
\label{eqn:d14}
\end{equation}
Let $m^{\prime}$ be the smallest integer satisfying $2\epsilon^{2} (m^{\prime}) /p = 2C_{1} \exp (- 2C_{2 }m^{\prime} ) /p < 1$.
If $m^{\prime} < m$,
$$\sum_{d^{8m-4}+1\leq j} \lambda_{j} \leq \exp [- 2C_{2}(m-m^{\prime})]. $$
Due to  (\ref{eqn:d13}) and (\ref{eqn:d14}), we obtain the following inequalities:
\begin{equation}
\sum_{j=1}^{d^{8m^{\prime}} -1} - \lambda_{j} \ln \lambda_{j} <  8m^{\prime} \ln d .
\label{eqn:d15}
\end{equation}
\begin{eqnarray}
&&\sum_{d^{8m-4} +1}^{d^{8m+4}} - \lambda_{j} \ln \lambda_{j} <  
(1-  \exp [- 2C_{2}]) \exp [- 2C_{2}(m-m^{\prime})] (8m-4)  \ln (D^{8}-1) 
\nonumber
\\
&& 
- \left(1-  \exp [- 2C_{2}]) \exp [- 2C_{2}(m-m^{\prime})] \{ \ln (1-  \exp [- 2C_{2}])  - 2C_{2}(m-m^{\prime}) \}\right) .
\nonumber
\\
&&
\label{eqn:d16}
\end{eqnarray}	
These estimates imply (\ref{eqn:d7}).
{\it End of Proof of Lemma \ref{lmm:lmm4.2}} 
\\
\\
{\it Proof of Proposition\ref{pro:pro4.2}}
\\
We fix a large number $S$ and suppose that $s(\varphi_{[j,i]}) > S$
for $i$. For any $k$ satisfying $k<i$, 
$$s(\varphi_{[ j , i]}) \leq s(\varphi_{[j,k]}) + s(\varphi_{[k+1,i]}) \leq s(\varphi_{[j,k]}) + (i-k)\ln d .$$
Setting $l_{0} = S/(3\ln d)$, we have $\frac{2}{3}S \leq s(\varphi_{[j,k]})$ for $k$ with $ i- l_{0} \leq k \leq i$ . 
Thus, if the entanglement entropy is not bounded, for any 
large $S_{cut}$ there exists $i$
\begin{equation}
S_{cut} \leq s(\varphi_{[- k, i+ k]}) 
\label{eqn:d17}
\end{equation}	
where $l_{0} = S_{cut}/(2\ln d)$ and $0 \leq k \leq l_{0}$.
\\
Due to Lemma \ref{lmm:lmm4.2},
\begin{equation}
p \leq 2C_{1}^{2} \exp [-(S_{cut} -F)/(C_{2}\ln 4d)]
\label{eqn:d18}
\end{equation}
Set $ x = \varphi_{L,i}\otimes\varphi_{R,i}(O_{B}(i,l))$ and 
$y= \varphi_{L,i}\otimes\varphi_{R,i}(O_{L}(i,l)O_{R}(i,l)) (\geq 1- 2\epsilon (l))$.
For any state $\psi$, any operators $E$,$B$ with $0\leq E,B\leq 1$,  the Schwartz inequality implies
\begin{eqnarray*}
|\psi ((E- \psi(E)1)(B - \psi(B)1))| &\leq& (\psi (E^{2}) - \psi (E)^{2} )^{1/2} (\psi (B^{2}) - \psi (B)^{2} )^{1/2}
\\
&\leq&  (\psi (E) - \psi (E)^{2} )^{1/2} (\psi (B) - \psi (B)^{2} )^{1/2}  .
\end{eqnarray*}
Setting $B= (O_{B}(i,l)$, $E = (O_{L}(i,l)O_{R}(i,l)$, $\psi = \varphi_{L,i}\otimes\varphi_{R,i}$
\begin{eqnarray*}
&&xy - | \varphi_{L,i}\otimes\varphi_{R,i}(O_{B}(i,l) O_{L}(i,l)O_{R}(i,l)) | \leq \sqrt{x-x^{2}} \sqrt{y-y^{2}} \: ,
\\
&& x(1-  2\epsilon (l)) - \sqrt{x} \sqrt{ 2\epsilon (l)} - \epsilon (l)
\\
&&\leq
 | \varphi_{L,i}\otimes \varphi_{R,i}(O_{B}(i,l) O_{L}(i,l)O_{R}(i,l)) | -  \epsilon (l) \leq p
\end{eqnarray*}  
\begin{equation} 
x\leq\left\{ 2C_{1}^{2} \exp [-(S_{cut} -F)/(C_{2}\ln 4d)] +\sqrt{x} \sqrt{ 2\epsilon (l)} + 2\epsilon (l)\right\}/
 (1- 2\epsilon (l))  
\label{eqn:d19}
\end{equation} 
We can find $C_{3}$ such that $x \leq C_{3} \epsilon (l) <1$.
We now assume that 
$C_{2}\ln C_{1} \ln 4d + F \leq  S_{cut}/2 $ and we obtain 
$$l \leq l_0 \leq  (S_{cut} -F)/(C_{2}\ln 4d) - C_{2} \ln C_{1} ,$$
$$2 C_{1}^{2} \exp [ (S_{cut} -F)/(C_{2}\ln 4d)] \leq 2 \epsilon (l) .$$
Due to (\ref{eqn:d19}),
\begin{equation}
x \leq \frac{4\epsilon (l) + \sqrt{ 2x\epsilon (l)} }{ 1- 2\epsilon (l)}.
\label{eqn:d20}
\end{equation}
which shows that $x \leq C_{4} \epsilon (l)$ for a constant $C_{4}$.
\par
On the other hand , due to monotonicity of relative entropy for states $\varphi_{[-l,l]\cup [i-l,i+l]}$ and 
$\varphi_{[-l,-1]\cup [i+1,i+l]}\otimes\varphi_{[0,l]\cup[i-l,i]}$
\begin{eqnarray}
&&(1- 2\epsilon (l)) \ln (1- 2\epsilon (l))/x  + 2\epsilon (l) \ln 2\epsilon (l)/(1-x) 
\nonumber\\
&&\leq \varphi_{[-l,l]\cup [i-l,i+l]} + s(\varphi_{[-l,-1]\cup [i+1,i+l]}) + s(\varphi_{[0,l]\cup[i-l,i]})
\label{eqn:d21}
\end{eqnarray}
(\ref{eqn:d21}) implies
$$- s(\varphi_{[-l,l]\cup [i-l,i+l]}) + s(\varphi_{[-l,-1]\cup [i+1,i+l]}) + s(\varphi_{[0,l]\cup[i-l,i]}) 
\geq (1- 2\epsilon (l)) \ln 1/x  -\ln 2 .$$
As a consequence we have a positive constant $C_{5}$ such that
$$- s(\varphi_{[-l,l]\cup [i-l,i+l]}) + s(\varphi_{[-l,-1]\cup [i+1,i+l]}) + s(\varphi_{[0,l]\cup[i-l,i]}) 
\geq (1- 2\epsilon (l)) \ln 1/\epsilon (l)  -C_{5}. $$ 
The above estimate is valid for $j,l$ if  $j+l \leq i+l_{0}$ and $l\leq l_{0}$:
\begin{equation}
- s(\varphi_{[-l,l]\cup [j-l,j+l]}) + s(\varphi_{[-l,-1]\cup [j+1,j+l]}) + s(\varphi_{[0,l]\cup[j-l,j]}) 
\geq (1- 2\epsilon (l)) \ln (1/\epsilon (l))  -C_{5}.
\label{eqn:d22}
\end{equation}
Suppose that $J$ and $K$ are any intervals of length $l$ in $[-l_{0}, i+l_{0}]$ and set
$$S_{l} = max\{ s(\varphi_{J\cup K})  \:\: | \:\: J,K \subset [-l_{0}, i+l_{0}].\: |K| = |J|=l \} .$$
By definition $S_{1}\leq \ln 2d$, and  due to (\ref{eqn:d22})  
$$ S_{2l} \leq 2 S_{l} - (1-2C_{1} \exp [- l/C_{2}]) l/C_{2} + \ln C_{1} + C_{5},$$
\begin{equation}
0\leq S_{2^{k}} \leq \ln 2d  2^{k}  - 2^{k} k /C_{2}  +C_{6} 2^{k}
\label{eqn:d23}
\end{equation}
where
$$ \sum_{m=0}^{\infty} 2C_{1} \exp [- 2^{m} /C_{2}]) l/C_{2} + \ln C_{1} + C_{5}$$
When we take $k$ sastisfying  $2^{k} \leq l_{0} < 2^{k+1}$,
\begin{equation}
\ln_{2} (S_{cut}/2 ) = \ln_{2} l_{0} \leq 1 + \ln 2d + C_{6}
\label{eqn:d24}
\end{equation}
Hence, we arrive at the contradiction to the claim that $S_{cut}$ can be an arbitrary large number.
\section{Spinless Fermion}\label{Fermion}
\setcounter{theorem}{0}
\setcounter{equation}{0}
In this section, we consider translationally invariant pure states of spinless Fermion systems on $\bfZ$.
Let us consider the GNS representation of $\frak A^{CAR}$
associated with a translationally invariant pure state $\psi$ and we show the fermionic
version of Haag duality.
In general, any translationally invariant factor state $\psi$ of  $\frak A^{CAR}$
is $\Theta$ invariant. (See \cite{ArakiMoriya} for basic properties of ferminonic systems.) 
Suppose that a state $\psi$ of  $\frak A^{CAR}$ is $\Theta$ invariant and 
let $\{\pi_{\psi} (\frak A_{CAR}), \Omega_{\psi}, \frak H_{\psi}\}$ be the GNS triple 
associated with $\psi$.
There exists a (unique) selfadjoint unitary $\Gamma$ on $\frak H_{\psi}$ 
satisfying
\begin{equation}
\Gamma \pi_{\psi}(Q)\Gamma^{-1}= \pi_{\psi}(\Theta(Q)) , \:\: 
\Gamma^{2}=1, \:\: \Gamma =\Gamma^{*}, \:\: \Gamma \Omega_{\psi} = \Omega_{\psi}. 
\label{eqn:z2}
\end{equation}
With aid of $\Gamma$, we introduce another representation $\overline{\pi}_{\psi}$
of $\frak A^{CAR}$ via the following equation:
\begin{equation}
\overline{\pi}_{\psi}(c_{j}) = \pi_{\psi} (c_{j})\Gamma, \:\: 
\overline{\pi}_{\psi}(c_{j}^{*}) = \Gamma \pi_{\psi} (c_{j}^{*})
\label{eqn:z3}
\end{equation}
for any integer $j$.
Let $\Lambda$ be a subset of $\bfZ$ and $\psi$ be a state of  $\frak A^{CAR}$ which is $\Theta$ invariant. 
By definition,$ \pi_{\psi}( \frak A^{CAR}_{\Lambda} )^{\dprime} \subset
\overline{\pi}_{\psi} ( \frak A^{CAR}_{\Lambda^{c}} )^{\prime}  $.
 We say the twisted Haag duality is valid for  $\Lambda$ if and only if
\begin{equation}
\pi_{\psi}( \frak A^{CAR}_{\Lambda} )^{\dprime} = 
\overline{\pi}_{\psi} ( \frak A^{CAR}_{\Lambda^{c}} )^{\prime} 
\label{eqn:z4}
\end{equation}
holds.  To formulate split property of fermion systems, we may consider
existence of an intermediate type I factor $\calN$ in
$ \pi_{\psi}( \frak A^{CAR}_{\Lambda} )^{\dprime} \subset \calN \subset
\overline{\pi}_{\psi} ( \frak A^{CAR}_{\Lambda^{c}} )^{\prime}  $.
We note that  $\pi_{\psi}( \frak A^{CAR}_{\Lambda} ) \cup
\overline{\pi}_{\psi} ( \frak A^{CAR}_{\Lambda^{c}} )$ may not act irreducibly on $\frak H_{\psi}$ 
even if $\psi$ is pure. It is possible to show the following.
\begin{lmm}
Supose that $\psi$ is a $\Theta$ invariant pure state of  $\frak A^{CAR}$ and consider the vector state
$\omega_{\psi}$ associated with $\Omega_{\psi}$ of $\pi_{\psi}( \frak A^{CAR}_{R} ) \cup
\overline{\pi}_{\psi} ( \frak A^{CAR}_{L} )$.
$\omega$ is not a pure state if and only if there exists a selfadjoint unitary $\Gamma_{-}$ satisfying
\begin{eqnarray}
&&\Gamma_{-} \pi_{\psi}(c_{j} ) = -  \pi_{\psi}(c_{j} )  \Gamma_{-} , \:
\Gamma_{-} \pi_{\psi}(c_{i} ) =  \pi_{\psi}(c_{i} )  \Gamma_{-}  \: ((j <0 ,0\leq i ), \:
\nonumber\\
&&\Gamma_{-}\Gamma = -\Gamma\Gamma_{-} .
\label{eqn:z400}
\end{eqnarray}
\label{lmm:FermiSpin1}
The commutant of $(\pi_{\psi}( \frak A^{CAR}_{R} ) \cup
\overline{\pi}_{\psi} ( \frak A^{CAR}_{L} ))^{\dprime}$ is generated by
$\Gamma_{-} $.
\end{lmm}
If we identify $\pi_{\psi}( \frak A^{CAR}_{R}$ with  $\pi_{\psi}( \frak A_{R} )$ and
$(\pi_{\psi}( \frak A^{CAR}_{L} )$ with $(\pi_{\psi}( \frak A_{L} )$  $\omega_{\psi}$ is a state
of $\frak A$. When $\psi$ is  $\Theta$ invariant pure state of  $\frak A^{CAR}$, we might say split property holds if 
$\pi_{\psi}( \frak A^{CAR}_{R})^{\dprime}$ is of type I.
We can show the following theorem in the same way as for the spin systems.  
\begin{theorem}
Let $\psi$ be a translationally invariant pure state of the CAR algebra
$\frak A^{CAR}$. and let $\{\pi_{\psi} (\frak A^{CAR}), \Omega_{\psi} , {\frak H}_{\psi}  \}$ 
be the GNS triple for  $\psi$. Suppose that  the entropy $s(\psi_{[1,L]})$ is bounded uniformly in $L$.
Consider the vector state $\omega$ assoicated with  
$ \Omega_{\psi}$ of  $\frak A^{CAR}_{(-\infty ,0]}\otimes\frak A^{CAR}_{[1,\infty )}$.
$\omega$ satisfies the split propery for $\frak A_{R}$ and $\frak A_{L}$. As a consequence,
$\pi_{\psi}( \frak A^{CAR}_{R})^{\dprime}$ is of type I and
the twisted Haag duality holds for $\Lambda =[1,\infty )$, 
$\pi_{\psi}((\frak A^{CAR})_L)^{\dprime} = 
\overline{\pi}_{\psi} ((\frak A^{CAR})_{R} )^{\prime} $.
\label{th:dualityFermi}
\end{theorem}
To show Theorem\ref{th:mainFermi2}, we employ the Jordan-Wigner transform for infinite systems 
\`a la mani\`ere de \cite{XY1},  \cite{XY2} .
Fermion systems and quantum spin chains are formally equivalent via the Jordan-Wigner 
For handling infinite chains, we introduce an automorphism $\Theta_{-}$ of $\frak A_{CAR}$ by the following equations: 
$$ \Theta_{-}(c^{*}_{j})= - c^{*}_{j} ,\:  \Theta_{-}(c_{j})= - c_{j}  \: ( j\leq 0), $$
$$ \Theta_{-}(c^{*}_{k})=  c^{*}_{k} ,\:  \Theta_{-}(c_{k})= c_{k} \: ( k >0 ) .$$
Let $\tilde{\frak A}$ be the crossed product of $\frak A_{CAR}$ by  the $\bfZ_{2}$ action $\Theta_{-}$ . 
$\tilde{\frak A}$ is the $C^{*}$-algebra generated by
$\frak A_{CAR}$ and a unitary $T$ satisfying 
$$ T=T^{*}, \: T^{2}= 1 , \: \: T Q T = \Theta_{-}(Q)\:\:\: ( Q \in \frak A_{CAR}).$$
Via the following formulae, we regard $\frak A$ as a subalgebra of $\tilde{\frak A}$:
\begin{align}
\sigma_z^{(j)}&=2c_j^*c_j -1
\nonumber\\
\sigma_x^{(j)}&=TS_{j}(c_j + c_j^*)
\nonumber\\
\sigma_y^{(j)}&=iTS_{j} (c_j - c_j^*).
\label{eqn:z6}
\end{align}
where
\begin{align*} 
S_{n}
=\left\{\begin{gathered}
\sigma_z^{(1)}\cdots\sigma_z^{(n-1)}\quad n>1\\
1\quad\quad\quad\quad n=1\\
\sigma_z^{(0)}\cdots\sigma_z^{(n)}\quad n<1.
\end{gathered}
\right.
\end{align*}
We extend the automorphism $\Theta$ of $\frak A_{CAR}$ to $\tilde{\frak A}$
via the  following equations:
$$ \Theta (T) = T  , \: \Theta ( \sigma_x^{(j)}) = - \sigma_x^{(j)} ,  \:
\Theta ( \sigma_y^{(j)}) = - \sigma_y^{(j)} , \: \Theta ( \sigma_z^{(j)}) =  \sigma_z^{(j)} .$$
As is the case of the CAR algebra, we set
$$(\frak A )_{\pm} = \{ Q \in \frak A |  \Theta (Q) = \pm Q \}, \:  
(\frak A_{\Lambda})_{\pm} = (\frak A)_{\pm} \cap \frak A_{\Lambda} , \:
(\frak A_{loc})_{\pm} = (\frak A)_{\pm} \cap \frak A_{loc} .$$ 
Then, it is easy to see that 
$$(\frak A)_{+} = (\frak A^{CAR})_{+} , \:  
(\frak A_{\Lambda})_{+} = (\frak A^{CAR}_{\Lambda})_{+} , \:
(\frak A_{loc})_{+} = (\frak A^{CAR}_{loc})_{+}  .$$ 
Let $\psi$ be a pure state of $\frak A_{CAR}$ and assume that $\psi$ is $\Theta$ invariant.
Let $\psi_{+}$ be the restriction of $\psi$ to  $ (\frak A^{CAR})_{+}=(\frak A)_{+} $.
$\psi_{+}$ is extendible to a $\Theta$ invariant state $\varphi_{0}$ of  $\frak A$
via the following formula:
\begin{equation}
 \varphi_{0} (Q)  =  \psi_{+}(Q_{+}), \quad  Q_{\pm} = \frac{1}{2} (Q\pm\Theta (Q)) \in (\frak A)_{\pm} .
\label{eqn:z7}
\end{equation}
In general,  $\varphi_{0}$  may not be a pure state but  if $\varphi$ is a pure state extension of  
$\psi_{+}$ to $\frak A$, the relation between $ \varphi_{0}$ and $\varphi$ is written as
$ \varphi_{0}(Q) =  \varphi (Q_{+}) $.
That $ \varphi_{0}$ and $\varphi$ are identical or not depends on existence of a unitary 
implementing $\Theta_{-}$ on $\frak H_{\psi}$. 
\begin{pro}
\label{pro:ext1}
Let $\psi$ be a $\Theta$ invariant pure state of $\frak A^{CAR}$ and $\psi_{+}$ be the restriction
of $\psi$ to  $(\frak A^{CAR})_{+}$.
\begin{description}
\item[(i)]
 Suppose that  $\psi$ and $\psi\circ\Theta_{-}$ are not unitarily equivalent.
The unique $\Theta$ invariant extension $\varphi$ of $\psi_{+}$ to $\frak A$ is a pure state. 
If $\psi$ is translationally invariant, $\varphi$ is translationally invariant as well.
\item[(ii)] 
 Suppose that  $\psi$ and $\psi\circ\Theta_{-}$ are unitarily equivalent and
that  $\psi_{+}$ and $\psi_{+}\circ\Theta_{-}$ are unitarily equivalent as states of
$(\frak A^{CAR})_{+}$.
The unique $\Theta$ invariant extension $\varphi$ of $\psi_{+}$ to $\frak A$ is a pure state. 
If $\psi$ is translationally invariant, $\varphi$ is translationally invariant as well.
\item[(iii)] 
 Suppose that  $\psi$ and $\psi\circ\Theta_{-}$ are unitarily equivalent and
that  $\psi_{+}$ and $\psi_{+}\circ\Theta_{-}$ are not unitarily equivalent as states of
$(\frak A^{CAR})_{+}$.
There exists  a pure state extension $\varphi$ of $\psi_{+}$ to $\frak A$ which is not
 $\Theta$ invariant. Furthermore,
 we can identify the GNS Hilbert spaces $\frak H_{\psi_{+}}$ and $\frak H_{\varphi}$ and
 \begin{equation}
 \pi_{\varphi}(\frak A)^{\dprime} =  \pi_{\varphi}((\frak A)_{+})^{\dprime}.
\label{eqn:z8}
 \end{equation}
If $\psi$ is translationally invariant, $\varphi$ is a periodic state with period 2,
$\varphi\circ\tau_{2} = \varphi$ and 
 \begin{equation}
 \pi_{\varphi}(\frak A_{L})^{\dprime} =  \pi_{\varphi}((\frak A_{L})_{+})^{\dprime}, \quad
  \pi_{\varphi}(\frak A_{R})^{\dprime} =  \pi_{\varphi}((\frak A_{R})_{+})^{\dprime}
 \label{eqn:z9}
 \end{equation}
where we set $(\frak A_{L,R})_{+} = (\frak A_{L,R}) \cap (\frak A)_{+}$.
\end{description}
\label{pro:duality21}
\end{pro}
{\it Proof of Proposition \ref{pro:duality21}}
\\
Set $X_{j} = c_{j} + c_{j}^{*}$. 
As $\psi$ is $\Theta$ invariant, the GNS space $\frak H_{\psi}$ is a direct sum of
$\frak H_{(\psi )}^{( \pm)}$ where
$$\frak H_{\psi}^{(+)} = \overline{ \pi_{\psi}((\frak A)_{+})\Omega} ,\quad         
\frak H_{\psi}^{(-)} = \overline{ \pi_{\psi}((\frak A)_{+} X_{j})\Omega}  .$$
The representation $\pi_{\psi}((\frak A)_{+})$  of $(\frak A)_{+}$ on $\frak H_{\psi}$  is 
decomposed into mutually disjoint irreducible representations on $\frak H_{\psi }^{(\pm)}$.
\par
Let $\psi$ and $\tilde{\psi}$ be $\Theta$ invariant states of ${\frak A}^{CAR}$. 
The argument in 2.8 of \cite{Voiculescu} shows that if $\psi_{+}$ and $\tilde{\psi}_{+}$ of
$(\frak A)_{+}$ are equivalent, $\psi$ and $\tilde{\psi}$  are equivalent.
Now we show (i). If pure states $\psi$ and  $\psi\circ\Theta_{-}$ are not equivalent,
 $\psi_{+}=\varphi_{+}$ is not equivalent to $(\varphi\circ\Theta_{-})_{+}$ and 
 $(\varphi\circ\Theta_{-}\circ Ad(X_{j}))_{+}$.
 Consider the GNS representation $\{\pi_{\varphi}(\frak A),\Omega_{\varphi}, \frak H_{\varphi}\}$
 of $\frak A$. If we restrict $\pi_{\varphi}$ to $(\frak A)_{+}$ it is the direct sum of two
 irreducible GNS representations associated with $\psi_{+}=\varphi_{+}$ and $(\varphi\circ\Theta_{-}\circ Ad(X_{j}))_{+}$.
 So we set 
 $$\frak H= \frak H_{\varphi} , \: \frak H= \frak H_{1}\oplus \frak H_{2}, \:
 \frak H_{1}= \frak H_{\varphi_{+}}, \:
  \frak H_{2}= \frak H_{(\varphi\circ\Theta_{-}\circ Ad(X_{j}))_{+}} .$$
 Any bounded operator $A$ on $\frak H$ is written in a matrix form,
 \begin{equation}
 A = \left(\begin{array}{cc}a_{11} & a_{12} \\a_{21} & a_{22}\end{array}\right)
 \label{eqn:z12}
 \end{equation}
 where $a_{11} $(resp. $a_{22} $) is a bounded operator on $\frak H_{1}$ (resp. $\frak H_{2}$)
 and $a_{12} $(resp. $a_{21} $) is a bounded operator from $\frak H_{2}$ to $\frak H_{1}$
 (resp. a bounded operator from $\frak H_{1}$ to $\frak H_{2}$.
 As $\psi_{+}=\varphi_{+}$ is not equivalent to $(\varphi\circ\Theta_{-}\circ Ad(X_{j}))_{+}$,
  \begin{equation}
 P = \left(\begin{array}{cc} a & 0 \\ 0 &b \end{array}\right) 
 \label{eqn:z13}
 \end{equation} 
is an element of $\pi_{\varphi} ((\frak A)_{+} )^{\dprime}$ and 
$\pi_{\varphi}(\sigma_{x}^{(j)})$ looks like
\begin{equation}
\pi_{\varphi}(\sigma_{x}^{(j)} ) = \left(\begin{array}{cc} 0 & d \\ d^{*} &0 \end{array}\right) 
\label{eqn:z14}
 \end{equation}
A direct computation shows that an operator $A$ of the matrix form  (\ref{eqn:z12}) 
commuting with (\ref{eqn:z13}) and  (\ref{eqn:z14}) is trivial. This shows that the state $\varphi$ is pure.
\\
The translational invariance of $\varphi$ follows from  translational invariance of $\psi$
and $\varphi (Q) = \psi (Q_{+})$.
 \par
(ii) of Proposition \ref{pro:duality21} can be proved by constructing the representation
of $\frak A$ on the GNS space of Fermion.
By our assumption, $\pi_{\psi_{+}}((\frak A)_{+})$ is not equivalent to
$\pi_{\psi_{+}}(Ad(X_{j}) (\frak A)_{+})$. Hence $\pi_{\psi_{+}}((\frak A)_{+})$  is equivalent
to $\pi_{\psi_{+}}(\Theta_{-}(\frak A)_{+}))$ and $\pi_{\psi_{+}}(Ad(X_{j})(\frak A)_{+}))$  is equivalent
to $\pi_{\psi_{+}}(\Theta_{-}(Ad(X_{j})\frak A)_{+}))$. It turns out that 
there exists a selfadjoint unitary $U(\Theta_{-})$ 
($U(\Theta_{-})^{*}=U(\Theta_{-})$, $U(\Theta_{-})^{2}=1$)  
on $\frak H_{\psi}$ such that
\begin{equation}
U(\Theta_{-}) \pi_{\psi}(Q) U(\Theta_{-})^{*},  \quad
U(\Theta_{-}) \in \pi_{\psi}((\frak A)_{+})^{\dprime}
\label{eqn:z15}
 \end{equation}
for any $Q$ in $\frak A^{CAR}$.
Any element $R$ of $\frak A$ is writtten in terms of fermion operators and $T$ as follows:
\begin{equation}
R =  R_{+} + TR_{-}   , 
 \label{eqn:z16}
 \end{equation}
where
$$ R_{+} = \frac{1}{2} ( R + \Theta (R)) \in (\frak A^{CAR})_{+} , \quad
 R_{-} = \frac{1}{2} ( TR - T\Theta (R)) \in (\frak A^{CAR})_{-} .$$
Using this formula, for any $R$ in $\frak A$, we set
\begin{equation}
\pi (R) =   \pi_{\psi} (R_{+}) + U(\Theta_{-})   \pi_{\psi} (R_{-} )
 \label{eqn:z17}
 \end{equation}
$\pi (R)$ gives rise to a representation of $\frak A$ on $\frak H_{\psi}$ and we set
\begin{equation}
\varphi (R) =  \left( \Omega_{\psi},   \pi (R) \Omega_{\psi}\right)  .
 \label{eqn:z18}
 \end{equation}
The representation $\pi (\frak A)$ is irreducible because
$\pi ((\frak A)_{+})^{\dprime}$ contains  $U(\Theta_{-})$ and hence $\pi (\frak A)^{\dprime}$
contains $\pi ((\frak A^{CAR})_{-})$ and  $\pi (\frak A)^{\dprime}= \frak B (\frak H_{\varphi})$.
\\
As in (i), the translational invariance of $\varphi$ follows  from $\Theta$ invariance of $\varphi$
(by construction) and translational invariance of $\psi$ .
\par
To show (iii), we construct an irreducible representation of $\frak A$ on 
the GNS space $\frak H_{+} =\overline{ \pi_{\psi_{+}}((\frak A^{CAR})_{+} )\Omega_{\psi}}$.
Now under our assumption there exists a selfadjoint unitary $V(\Theta_{-})$ satisfying
\begin{equation}
V(\Theta_{-}) \pi_{\psi}(Q) V(\Theta_{-})^{*} = \pi_{\psi}(\Theta_(Q)),  \quad
 V(\Theta_{-}) \in \overline{\pi_{\psi}((\frak A)_{-})}^{w}
\label{eqn:z19}
 \end{equation}
for any $Q$ in $\frak A^{CAR}$. For $R$ written in the form (\ref{eqn:z16}), we set 
\begin{equation}
\pi (R) =   \pi_{\psi} (R_{+}) + V(\Theta_{-})   \pi_{\psi} (R_{-} )
 \label{eqn:z20}
 \end{equation}
for $R$ in $\frak A$ and $\pi (R)$ belongs to the even part $\pi_{\psi}((\frak A^{CAR})_{+})^{\dprime}$.
and $\pi (\frak A)$ acts irreducibly on $\frak H_{+}$.
\\
To show periodicity of the state $\varphi$, we introduce a unitary $W$
satisfying
$$W\Omega_{\psi}=\Omega_{\psi} , \quad 
W \pi_{\psi} (Q) W^{*} =  \pi_{\psi} (\tau_{1}(Q))  , \quad Q \in \frak A^{CAR}$$
The adjoint action of both unitaries $W V(\Theta_{-}) W^{*}$ and 
$V(\Theta_{-}) \pi_{\psi}(\sigma_{z}^{(1)})$ gives rise to the same automorphism on 
$\pi_{\psi}(\frak A^{CAR})$. 
By irreducibility of the representation $\pi_{\psi}(\frak A^{CAR})$, 
$W V(\Theta_{-}) W^{*}$ and $V(\Theta_{-}) \pi_{\psi}(\sigma_{z}^{(1)})$ 
differ in a phase factor.
\begin{equation}
W V(\Theta_{-}) W^{*} = c V(\Theta_{-}) \pi_{\psi}(\sigma_{z}^{(1)}) 
 \label{eqn:z21}
 \end{equation}
where $c$ is a complex number with $|c| =1$ .
 As both sides in (\ref{eqn:z21}) are selfadjoint , $c = \pm 1$.
Then, 
\begin{equation}
W^{2} V(\Theta_{-}) (W^{2})^{*} =  V(\Theta_{-}) \pi_{\psi}(\sigma_{z}^{(1)}\sigma_{z}^{(2)})
 \label{eqn:z22}
 \end{equation}
 This implies that the state $\varphi$ is periodic, $\varphi\circ\tau_{2}=\varphi$. 
 \\
 {\it End of Proof of Proposition \ref{pro:duality21}}
 
\begin{rmk}
In \cite{XXZ}, using expansion technique(but not the exact solution) 
we have shown  the XXZ Hamiltonian $H_{XXZ}$ with large Ising type anisotorpy 
$\Delta >>1$ 
$$H_{XXZ} = \sum_{j=-\infty}^{\infty} \{ \Delta \sigma_{z}^{(j)}\sigma_{z}^{(j+1)} +
\sigma_{x}^{(j)}\sigma_{x}^{(j+1)} + \sigma_{y}^{(j)}\sigma_{y}^{(j+1)} \}$$
has exactly two pure ground states 
$\varphi$ and 
$$\varphi\circ\Theta=\varphi\circ\tau_{1} \ne \varphi .$$ 
The unique $\Theta$ invariant ground state  $(1/2\varphi +\varphi\circ\tau_1)$ is 
a pure state of $(\frak A)_{+}$
In this example, the phase factor $c$ of (\ref{eqn:z21}) is $-1$. 
\end{rmk}
\bigskip
\bigskip
\par
To complete our proof of Theorem\ref{th:mainFermi2} , we use a theorem of \cite{Split} and 
Proposition \ref{pro:U(1)2} below.
\begin{theorem}
\label{th:Matsui2001}
Suppose that the spin $S$ of one site algebra $M_{2S+1}$($n=2S+1$) for $\frak A$ is
$1/2$.
Let $\varphi$ be a translationally invariant pure state of $\frak A$ such that
$\varphi_{R}$ gives rise to a type $I$ representation of $\frak A_{R}$.
Suppose further that  $\varphi$ is $U(1)$ gauge invariant , $\varphi\circ\gamma_{\theta}=\varphi$.
Then,  $\varphi$ is a product state.
\end{theorem}
\begin{pro}
\label{pro:U(1)2}
Let $\psi$ be a translationally invariant pure state of $\frak A^{CAR}$. 
\\
(i) Suppose further that  $\psi$ is $U(1)$ gauge invariant, $\psi\circ\gamma_{\theta}=\psi$.
The $\Theta$ invariant extension of $\psi_{+}$ to $\frak A$ is a translationally invariant pure
state.
\\
(ii) Suppose the conditions of (i) and that the von Neumann algebra
 $\pi_{\psi}(\frak A^{CAR}_{L})^{\dprime}$ associated with the GNS representation of
 $\psi_{L}$ is of type $I$. 
 Then,  either $\psi = \psi_{F}$  or $\psi = \psi_{AF}$ holds.
\end{pro}
\bigskip
\noindent
\newline
 {\it Proof of Proposition \ref{pro:U(1)2}}
\\
To prove Proposition \ref{pro:U(1)2} (i), we show
the case (iii) in Proposition \ref{pro:ext1}  is impossible due to assumption of 
$\gamma_{\theta}$ invariance.
There exists $U(\theta)$ implementing $\gamma_{\theta}$ 
on the GNS space of $\psi$. Then
$$U(\theta)  V(\Theta_{-}) U(\theta)^{*} = c(\theta) V(\Theta_{-}) $$
as the adjoint action of both unitaries are identical. Moreover these are selfadjoint 
so   $c(\theta)=\pm 1$ . Due to continuity in $\theta$ we conclude that $c(\theta)=1$
and $V(\Theta_{-})$ is an even element.
\par
Finally, we consider Proposition \ref{pro:U(1)2} (ii).  Due to (i) of Proposition \ref{pro:U(1)2} (i),
the Fermionic state $\psi$ has a translationally invariant pure state extension $\varphi$ to $\frak A$.
Then, the split property for Fermion implies that that of the Pauli spin system. It turns out that 
either $\psi (c^{*}_{j}c_{j}) =\varphi (e^{(j)}_{1}) = 0 $
or $\psi (c_{j}c_{j^{*}}) =\varphi (e^{(j)}_{2}) = 0$ holds. This completes our proof of Proposition 
\ref{pro:U(1)2} (ii). 
\\
 {\it End of Proof of Proposition \ref{pro:U(1)2}}
 \\
 \\
 If $\psi$ is a U(1) gauge invariant ground state with specral gap, the entanglement entropy
 is bounded and $\pi_{\psi}(\frak A^{F}_{R})^{\dprime}$ is of type I. Thus $\psi$ is trivial, which shows
 Theorem\ref{th:mainFermi2} .


\newpage
\begin{thebibliography}{99}

\bibitem{Araki1dim} Araki, H.:
{\it On Uniqueness of KMS states of One-dimensional Quantum Lattice Systems.}
 Comm. Math. Phys. {\bf 44}, 1--7(1975). 

\bibitem{XY1} Araki, H.:
{\it On the $XY$-model on two-sided infinite chain.}
 Publ. Res. Inst. Math. Sci.{\bf 20} no. 2, 277-296 (1984).

\bibitem{XY2} H.Araki, H., Matsui,Taku:
{\it Ground states of the $XY$-model.}
 Comm. Math. Phys. {\bf101 }, no. 2, 213--245(1985). 

\bibitem{ArakiMoriya} Araki,H.,  Moriya,H.:
Equilibrium statistical mechanics of Fermion lattice systems.
h Reviews in Mathematical Physics. {\bf15},  93-198.(2003)
 
 \bibitem{AffleckLieb}  Affleck, I. and Lieb, E.:
A proof of part of Haldane's conjecture on quantum spin chains,
Lett. Math. Phys. {\bf 12}, 57--69 (1986)


\bibitem{AKLT} Affleck, I., Kennedy, T., Lieb, E.H., Tasaki, H.: 
Valence Bond Ground States in Isotropic Quantum Antiferromagnets. Commun. Math. Phys.{\bf 115}, 477--528 (1988) 

\bibitem{BratteliRobinsonI} O.Bratteli and D.Robinson,
{\it Operator algebras and quantum statistical mechanics I} , 2nd edition
(Springer, 1987).

\bibitem{BratteliRobinsonII}O.Bratteli and D.Robinson,
{\it Operator algebras and quantum statistical mechanics II} , 2nd edition
(Springer, 1997).

\bibitem{DoplicherLongo} 
Doplicher,S., Longo,R.,:
 {\it Standard and split inclusions of von Neumann algebras.}  Invent.Math.{\bf 75},493-536(1984)

\bibitem{FNW} Fannes,M. , Nachtergaele,B. , Werner,R. :
{\it Finitely Correlated States on Quantum Spin Chains} , Commun.Math.Phys. {\bf 144},443-490 (1992). 

\bibitem{FNW2} Fannes,M. , Nachtergaele,B. , Werner,R. :
{\it Finitely correlated  pure states}
J.Funct. Anal. {\bf 120}, 511-534 (1994).

\bibitem{Nachtergaele2009}
Hamza, Eman; Michalakis, Spyridon; Nachtergaele, Bruno; Sims, Robert: 
Approximating the ground state of gapped quantum spin systems.  J. Math. Phys.  50  (2009),  no. 9, 095213, 16 pp.

\bibitem{Koma2006} Hastings, M.B. and Koma, T.:
Spectral Gap and Exponential Decay of Correlations,
Commun. Math. Phys. {\bf 265}, 781--804 (2006).

\bibitem{Hastings2007}
Hastings, M., "An area law for one dimensional quantum systems," J. Stat. Mech.: Theory Exp. 2007, P08024.

\bibitem{KMSW} Keyl,M., Taku Matsui, Schlingemann,D., Werner,R.F.,:
{\it Entanglement, Haag-Duality and Type Properties of Infinite Quantum Spin Chains.}
Rev.Math.Phys. {\bf 18},935-970(2006)

\bibitem{KMSW2}  Keyl,M.,Matsui,Taku,  Schlingemann,D., Werner,R.F.:
{\it On Haag-Duality  of Infinite Quantum Spin Chains.}
Rev.Math.Phys. {\bf 20},707-724(2008)

\bibitem{LiebRobinson} Lieb, E.H. and Robinson, D.W.:
The Finite Group Velocity of Quantum Spin Systems,
Commun. Math. Phys.  {\bf 28}, 251--257 (1972).

\bibitem{Longo1} Longo, R.:
{\it Solution to the factorial Stone-Weierstrass
 conjecture. An application of standard split $W^*$-inclusion.}  
Invent.Math. {\bf 76},145-155(1984)

\bibitem{TM1} Matsui,T. {\it A characterization of finitely correlated
pure states} 
{\it Infinite dimensional analysis and
quantum probability} Vol 1. , 647-661. 1998

\bibitem{Split} Matsui,Taku:
{\it The Split Property and the Symmetry Breaking of 
the Quantum Spin Chain.} 
Commun.Math.Phys.{\bf 218} , 393--416(2001). 

\bibitem{XXZ} Matsui,Taku:
{\it On the absence of non-periodic ground states for the antiferromagnetic $XXZ$ model. } 
Comm. Math. Phys. .{\bf 253} , 585--609(2005). 

\bibitem{Split2} 
Matsui, Taku: 
Spectral gap, and split property in quantum spin chains.  J. Math. Phys.  51  (2010),  no. 1, 015216, 8 pp. 82B20

\bibitem{Ruelle1962}
Ruelle, D.: {\it On the asymptotic condition in quantum field theory},
Helv. Phys. Acta {\bf 35}, 147 (1962).

\bibitem{nachtergaele2005}
Nachtergaele, B. and Sims, R.:
Lieb-Robinson Bounds and the Exponential Clustering Theorem,
Commun. Math. Phys. {\bf 265}, 119-130 (2006).

\bibitem{Nachtergaele2007a}
Nachtergaele, B. and Sims, R.:
Recent Progress in Quantum Spin Systems,
Markov Processes Relat. Fields {\bf 13}, 315-329 (2007).

\bibitem{NachtergaeleSims2009}  
Nachtergaele, B. and Sims, R.:
Locality Estimates for Quantum Spin Systems,
in New trends in mathematical physics. Selected contributions of the XVth international congress on mathematical physics. 
Springer (2009). 

\bibitem{Powers} Powers,R.T. {\it An index theory for semigroups of 
*-endomorphisms of $\calB ( \calH)$ and type $II_1$ factors}
Canad.J.Math.${\bf 40}$ (1988), 86-114.

\bibitem{Voiculescu} Stratila,S., Voiculescu,D.:
{\it On a Class of KMS States for the Unitary Group $U(\infty )$.}
Math.Ann. {\bf235},87-110(1978).

\bibitem{eisert2010}
Eisert,J., Cramer,M., and Plenio,M. B.:
{\it Area laws for the entanglement entropy}
Rev.Mod.Phys.82, 277?306 (2010) 

\bibitem{Hal2004}
Tasaki, H.: Low-lying excitation in one-dimensional lattice electron system. http://arXiv.org/list/cond-mat/0407616, 2004

\bibitem{Affleck1997}
Yamanaka, M., Oshikawa, M., Affleck, I.: 
Nonperturbative approach to Luttinger's theorem in one dimension. Phys. Rev. Lett.{\bf 79}, 1110--1113 (1997)

\end{thebibliography}
\end{document}